\documentclass[twocolumn,10pt,final,journal,oneside]{IEEEtran}
\usepackage{mathrsfs,amsfonts,amssymb,amsmath}
\usepackage{graphicx,cite,multirow}
\usepackage[linesnumbered, ruled]{algorithm2e}
\usepackage{siunitx}
\usepackage[skip=4pt]{caption}
\usepackage{subcaption}

\newcommand{\fref}[1]{Fig.~\ref{#1}}

\newcommand{\tref}[1]{Tab.~\ref{#1}}
\newcommand{\scref}[1]{Sec.~\ref{#1}}

\newcommand{\name}{UAVLoc\xspace}
\newcommand{\bm}[1]{\mbox{\boldmath{$#1$}}}

\begin{document}

\title{Edge-Assisted Multimodal UAV Localization with Resource-Efficient Compression and Robust Fusion}

\author{Zhong Ye, Yinghui He, Guanding Yu, \IEEEmembership{Senior Member,~IEEE}, and Pavel Loskot, \IEEEmembership{Senior Member,~IEEE}

\thanks{Manuscript received 7 February 2026; revised 2 July 2026, 
16 September 2026; accepted 5 October 2026.  
{\it(Corresponding authors: Yinghui He and Pavel Loskot)}
}

\thanks{Z. Ye, Y. He, and G. Yu are with the College of Information Science and Electronic Engineering, Zhejiang University, Hangzhou 310027, China (email: \{12431118, 2014hyh, yuguanding\}@zju.edu.cn).

P. Loskot is with the University of Illinois Urbana-Champaign Institute, Zhejiang University, Haining 314400, China (email: pavelloskot@intl.zju.edu.cn). }}

\markboth{IEEE TRANSACTIONS ON MOBILE COMPUTING, Vol. XX, No. X, Month 2026}{}

\maketitle

\begin{abstract}

  Unmanned aerial vehicles (UAVs) will play an important role in future urban transportation systems. This requires designing robust localization schemes especially for non-cooperative UAVs that do not share any information about their movements. This paper designs a multimodal UAV localization framework which utilizes camera, LiDAR, and radar sensing modalities. The underlying data processing and the subsequent inference of the UAV location are distributed among the sensing nodes and the edge server attached to the base station. The proposed UAV localization framework addresses three key challenges. First, the sensing nodes have limited computing and communication resources, and they may contain only single modality sensors. Second, the multimodal data differ greatly in sampling rates, time alignment and the encodings. Third, the changes in the environment and the hardware failures cause the modal data to degrade, or to be completely missing. The proposed localization framework utilizes several data processing modules including an information-bottleneck (IB)-based compression module that extracts the most relevant features from each modality, a time-encoding alignment module that provides the unified representations in a shared latent space, a multimodal fusion module that accounts for the degraded and missing data, and a Mamba-based regression module that predicts the present UAV location. The experiments involving a real-world dataset demonstrate that the proposed framework accurately and reliably obtains the UAV location while outperforming other existing frameworks.
\end{abstract}


\section{Introduction}

Up-to-date traffic information is crucial in managing the urban traffic flows, for example, to mitigate the congestion during peak hours and during accidents. However, the real-time traffic monitoring is challenging, which creates many inefficiencies, and increases the costs, especially in the last-mile delivery segments \cite{dui2024iot, garg2023drones, ye2024isac}. The recent rapid progress in the development of unmanned aerial vehicles (UAVs) provides new opportunities for traffic monitoring by exploiting greatly improved visibility of traffic situations from low altitudes above the ground. The UAVs can also be used as temporary communication relays over complex terrains by providing the line-of-sight connectivity \cite{telikani2025unmanned} as well as good sensing capabilities in urban environments \cite{banafaa2024comprehensive, debnath2024review, hawashin2024blockchain}.

The large-scale UAV deployment in urban areas faces significant challenges in low-altitude airspace management. The low-altitude airspace is structurally complex, and involves diverse flight characteristics and behaviors \cite{yuan2024mmaud, zhao2022vision}. In order to ensure the safe and orderly operations, the ground-based supervisory systems must continuously and accurately localize the UAVs in addition to their self-reported positions, since the latter may be accidentally or intentionally misleading, or missing. Accurate tracking of the UAVs is also important in the downstream supervisory tasks including the trajectory analysis, and the airspace occupancy assessment.

The external sensing of the UAV locations is essential for monitoring non-cooperative UAVs, when the data from their onboard navigation equipment are not available. Since different modalities are affected by the environmental conditions differently, the single-modality methods cannot provide reliable localization estimates in all scenarios. In particular, the vision cameras are affected by illumination changes and occlusion. The LiDAR has limited resolution at larger distances. The millimeter-wave (mmWave) radar suffers from a low angular resolution and multipath distortion \cite{couturier2024review, khawaja2025survey, seidaliyeva2025lidar}. Thus, combining multiple sensing modalities can provide a robust solution \cite{ye2025radar}. The existing multimodal methods rely on the data provided by the UAV's onboard navigation equipment \cite{gyagenda2022review, tong2023multi}. This makes these methods inherently unsuitable when the UAVs are non-cooperative. The recent studies also explored the UAV positioning and multimodal UAV communications. The study \cite{xu2025transformer} explored a FAS-enabled cooperative 3D UAV positioning by jointly optimizing the UAV trajectories and the antenna-port selection. The work \cite{xin2024novel} proposed a CNN-transformer based multimodal fusion for predicting the UAV-to-ground communication channels.

This paper investigates localization of a non-cooperative UAV, and addresses the corresponding data pre-processing, edge-side compression, and robust multimodal fusion in the resource-constrained scenarios. In particular, a novel, purely external multimodal UAV localization framework is devised, which does not require any data to be actively reported from the UAVs. The multimodal sensing is provided by the ground-placed cameras, LiDAR, and radar. The multimodal measurements are fused at the edge server of the cellular base station. A practical deployment of such a framework needs to overcome several challenges. Specifically, the network edge servers have limited computational and communication resources, so they cannot easily accommodate another functionality. The wireless links may not have sufficient capacity to guarantee a bounded transmission latency for time-sensitive multimodal data \cite{lahmeri2021artificial, sun2024advancing}. Moreover, the multimodal data are highly heterogeneous, and they differ in sampling rates, timestamps, and dimensionality, while their semantic and geometric representations often vary substantially. The environment disturbances and the limited sensing ranges can cause certain modalities to degrade, or to become temporarily unavailable. This leads to instability in the modal fusion, and the subsequent inference of the UAV location \cite{ma2021smil, liang2024foundations}. Other issues arise in performing the multimodal fusion itself. For example, the transformer-based fusion schemes utilizing cross-attention are sensitive to data scaling and noise \cite{tsai2019multimodal, chen2024transformer}, which make them unsuitable in the practical scenarios considered in this paper.

The following four techniques were developed to overcome or at least to mitigate the above-mentioned issues. First, the Information-Bottleneck Collaborative Compression (IB-CC) is proposed to compress the modal features into compact latent representations. This technique is inspired by the IB principle \cite{tishby2000information, hu2024survey}. Thus, only information that is most relevant to the localization task is retained, which saves both communication and computing resources. All modal data are projected into a shared latent space to facilitate the subsequent multimodal fusion. Second, the TE-Align module is introduced to cope with inconsistent sampling rates and timestamp variations across different modalities. It uses a unified timestamp encoding and a window-based feature aggregation to map the heterogeneous observations into a common window-level temporal representation. Third, the Mask-Aware Dynamic Weighting (MADW) module adaptively adjusts the modality contributions based on their task-dependent confidence scores and modal data availability, which improves the fusion robustness when the modal data are degraded or absent. Fourth, the Dependency-aware Cross-Attention (DACA) module enhances robustness of the feature fusion by learning the nonlinear task-relevant cross-modal dependencies. It allows focusing on the informative latent features, while reducing the influence of noisy and weakly correlated signals.

The main contributions of this paper can be summarized as follows.
\begin{itemize}
  \item A multimodal UAV localization framework is devised, which can be deployed on the resource-constrained devices at the network edge. The feature extraction and the data compression are integrated at the sensing nodes, whereas the multimodal alignment, modal fusion, and localization inference are performed on the edge server connected to a base station.
  \item The four new techniques, namely IB-CC, TE-Align, MADW, and DACA are developed to overcome the practical implementation challenges including the limited computing and communication resources, strong modal heterogeneity, and the problem with temporarily degraded and missing modalities.
  \item Extensive numerical experiments were conducted involving the publicly available real-world dataset to demonstrate the effectiveness and the superior performance of the proposed framework over other representative baseline schemes.
\end{itemize}

The remainder of this paper is organized as follows. \scref{sc:2} reviews the relevant literature. The system architecture is described in \scref{sc:3}. The modality-specific data processing is introduced in \scref{sc:4}. The proposed framework is presented in \scref{sc:5}. Numerical results are reported and evaluated in \scref{sc:6}. The paper is concluded in \scref{sc:7}.

\section{Related Work} \label{sc:2}

The proposed UAV localization framework assumes the camera vision, LiDAR and radar modalities. The multimodal data fusion and the subsequent location inference need to be designed to achieve the computational and transmission efficiency as well as robustness in typical deployment scenarios.

Camera vision is probably the most widely used modality for UAV localization, since these sensing devices are relatively cheap, easy to maintain, and provide rich information about the target. The visual localization of UAVs in global navigation satellite system (GNSS)-denied environments was considered in \cite{ye2024coarse} and \cite{xu2024visual}. The former achieved the decimeter-level accuracy using a coarse-to-fine data transformation. The latter integrated the visual odometry and the image-based geolocation for the real-time pose estimation. The computational complexity of visual UAV localization was reduced in \cite{wang2025lightweight} by adopting a lightweight deep-learning approach.

The LiDAR can be effective for the UAV detection and tracking by exploiting its high-resolution 3D sensing capabilities. The voxel mapping of LiDAR images with a cluster-based multi-target tracking for fast relative localization of UAVs was developed in \cite{vrba2024onboard}. An unsupervised LiDAR-based detection and tracking method for reliable target localization in the real-world scenes was proposed in \cite{liang2025unsupervised}. This scheme fuses multiple space-time scans, and it does not require labeled data for training.

Unlike other methods, the radar can be effective in adverse weather and poor visibility conditions. It excels at detecting the targets with sufficient reflectivity, which can be satisfied for some UAVs. For example, the work \cite{wen20242d} proposed a method for 3D localization from the 2D angular measurements provided by a monostatic MIMO radar. The convolutional neural network (CNN) for processing the range-Doppler maps was considered in \cite{tian2024fully} to enable the fast UAV detection and tracking with high accuracy. The mmWave radar system was designed in \cite{abdelnasser2024radro} to provide fast 3D localization in the cluttered indoor environments without relying on any anchors.

The single-modality schemes can achieve a high accuracy in some environments, however, they all have their inherent limitations. The multimodal schemes were proposed to exploit the complementary characteristics of different sensing modalities, which make them better suited to dynamic environments. The fusion of radio-frequency (RF) sensing and optical vision was considered in \cite{xie2024rf, 11288062}. It jointly calibrates the antenna array and the camera in order to align the RF signals with the image regions. It is particularly effective for real-time localization of small UAVs. An asynchronous LiDAR and camera fusion framework was proposed in \cite{xu2024lcdl} for the robust outdoor trajectory estimation. Assuming the complex and dense urban environments, a low-cost UAV self-localization method was developed in \cite{cui20253d}. It fuses the camera-inferred depths with a point cloud from the clustered mmWave radar using a geometric image registration to achieve the decimeter-level accuracy. The graph neural networks (GNNs) for indoor localization have been investigated in \cite{he2025sensem, xiao2025meta}. The multimodal approaches proposed in the literature are often optimized for the specific combinations of sensors and environments under idealized conditions including stable sensing and reliable data transfers. The effectiveness of AI-assisted localization systems also depends on their integration within specific application scenarios \cite{jin2026profiles}.

This paper addresses some of the practical issues that are rarely considered in the literature. The goal is to devise a UAV localization scheme that is not only accurate, but also robust under diverse sensing conditions. In particular, the embedded devices have constrained communication and computing resources, whereas the multimodal sensors continuously generate large volumes of raw data at different frame rates. Offloading the raw data to the edge server is difficult due to the bandwidth constraints, elevated energy consumption, and the induced transmission delays. This makes real-time localization impossible to accomplish at the embedded devices in sensing nodes. Another issue is time alignment of heterogeneous multimodal data that have different frame rates, dimensionality, and the fields of view. For example, the cameras provide high-resolution imagery, but with limited spatial perception. The 3D point cloud generated by LiDAR has slow refresh cycles. The radar has quick response to target motions, but it does not capture the environment details. The fusion of these multimodal data is challenging in order to avoid the cross-modal noise amplification, and alignment interference. Practical sensing is also affected by changes in the environment, and by the limited sensing ranges. Moreover, sensing device failures can degrade a sensing modality or even cause its complete loss. Resolving these issues, for example, by providing an adaptive fusion mechanism, is critical for achieving robust sensing to enable reliable
UAV localization in the downstream.

\section{System Architecture} \label{sc:3}

\begin{figure}[t]
  \centering
  \includegraphics[width=0.9\linewidth]{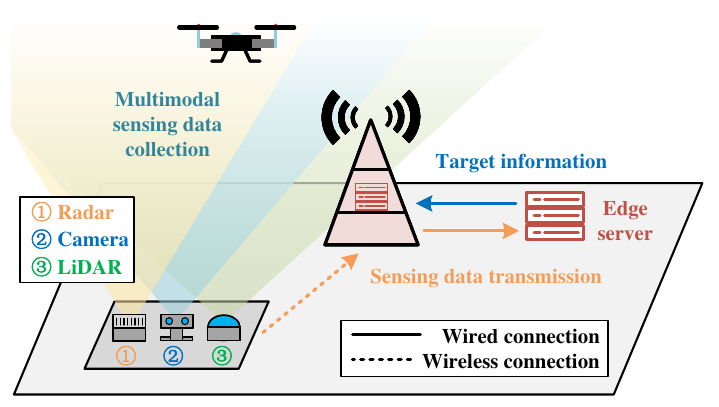}
  \caption{A distributed UAV localization scheme involving different sensing modalities.}
  \label{fig:model}
\end{figure}

Consider a multimodal UAV localization scenario depicted in \fref{fig:model}. It incorporates multiple sensing modalities such as binocular cameras, radar, and LiDAR, which continuously scan the nearby airspace. The task is to locate the position and the trajectory of a non-cooperative UAV, when it does not share any information about its movements. The large volumes of continually produced sensing data require that the raw data are pre-processed locally in order to reduce the communication bandwidth requirements under strict latency constraints. It is assumed that the sensing nodes are equipped with embedded low-power computing devices such as Raspberry Pi \cite{fezari2023raspberry}. These devices enable the sensing data to be compressed and encoded by extracting the key features following the principle of information bottleneck (IB). The pre-processed data are then forwarded to a base station (BS), which offloads heavy computations to a connected edge server. The server performs the multimodal fusion and the UAV location estimation. In addition, we make the following assumptions: (i) the sensing nodes are time synchronized, and the timestamps of the collected sensing data are globally aligned, (ii) unreliable uplink transmissions may result in missing sensing data, and (iii) the buffering delay is assumed to be negligible relative to the duration of the adopted alignment window.

\begin{figure}[t]
  \centering
  \includegraphics[width=0.9\linewidth]{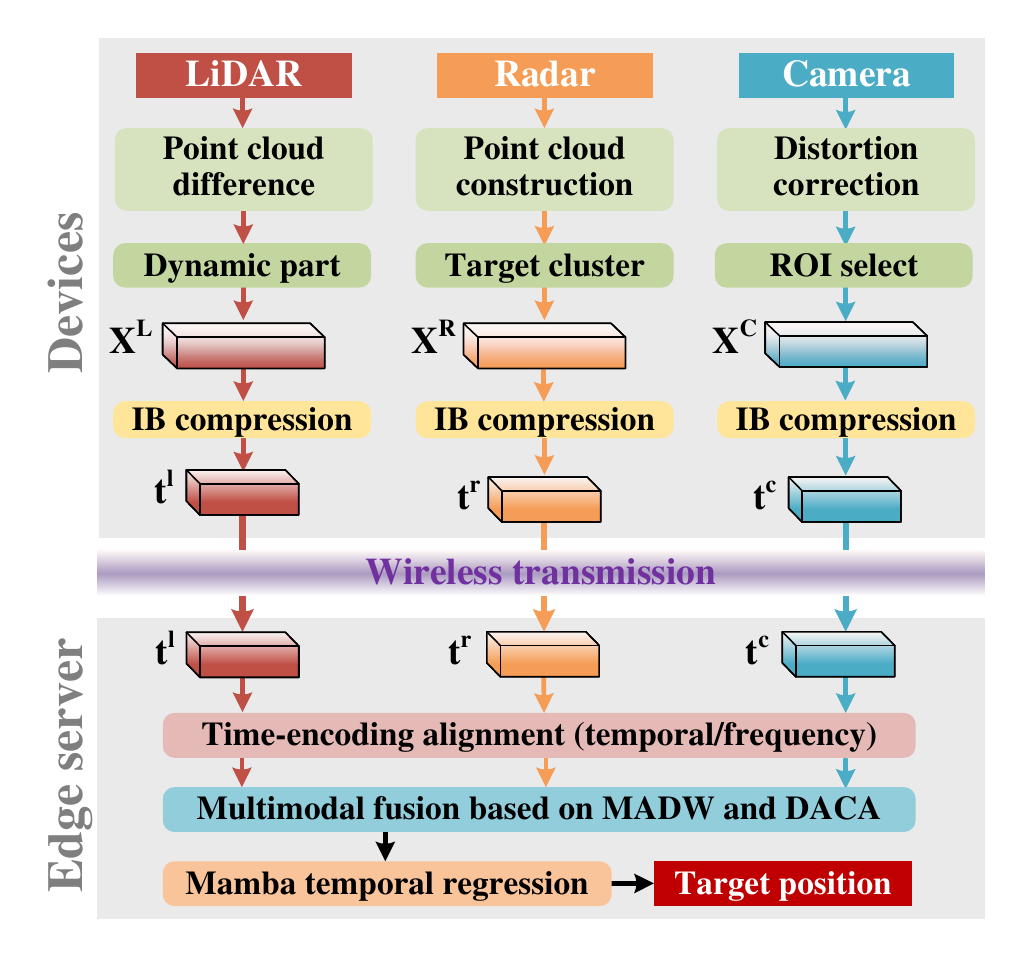}
  \caption{The architecture of the proposed \name, a multimodal UAV localization framework.}
  \label{fig:mib}
\end{figure}

The proposed collaborative multimodal UAV localization framework referred to as \name leverages multiple sensing modalities in order to provide an accurate localization of the fly-by UAV in real time. The localization is performed in two stages as indicated in \fref{fig:mib}. On the sensing side, each modality undergoes the modality-specific data pre-processing including extraction and compression of the important features before they are transmitted to the BS. At the server, the data forwarded from the BS are first decoded and temporally aligned before they are fused. The final step of the distributed data processing pipeline is estimation of the present location of the target UAV.

The sensing-side modality-specific processing can be summarized as follows. The fisheye distortion of the camera is rectified, and the regions of interest (ROI) are extracted for the subsequent processing. The point cloud from LiDAR undergoes a dynamic-static differentiation to extract the relevant dynamic features. The radar measurements are first converted into a point cloud. The data points are then clustered to select the points containing relevant information about the target. An IB-based compression scheme is designed to encode the extracted features for each modality. The IB encoders map the heterogeneous modal features into a shared latent space. The resulting latent representations are compact, which reduces the transmission requirements when uploading the data to the edge server via the BS.

The data processing at the edge server must account for heterogeneity of different modalities. In particular, the data of different modalities must be temporally aligned despite their different sampling rates and timestamps. The TE-Align module performs such an alignment by generating the unified multimodal representations. In addition, the MADW and DACA modules are used to stabilize the multimodal fusion, when the modal data are corrupted by noise to a different degree, or even completely missing. Specifically, the MADW module learns modality-specific confidence scores from the aligned latent representations, and combines them with availability information of different modalities to obtain the adaptive fusion weights. The DACA module guides the cross-modal attention by exploiting a learnable dependency score between the modalities. The fused representations are fed into a Mamba-based temporal regression network. The Mamba-based modeling of temporal dependencies allows the model to obtain the stable estimates of the UAV's positions over longer-term horizons.

The proposed framework is designed for the resource-constrained sensing of nearby airspace and localization of a non-cooperative UAV by utilizing heterogeneous sensing modalities that are typically temporally misaligned, and potentially degraded or missing. This paper addresses two key aspects. First, each module is adapted to the constraints of the considered practical scenarios. In particular, the IB-CC module extracts task-relevant compact representations in order to reduce the communications overhead. The TE-Align module maps the heterogeneous asynchronous data streams into a common window-level temporal representation under the globally aligned timestamps. The time-aligned representations are used by the MADW module to mitigate the adverse effects of unreliable modalities. The DACA module exploits complementary cross-modal information beyond the conventional dot-product similarity. Second, these modules are integrated into a coordinated task-oriented data processing pipeline. Together with a Mamba-based temporal regression, they jointly address the challenges of the constrained communication and computing resources, the modal heterogeneity, and the problem of degraded or missing modalities, which enables robust multimodal UAV localization.

\section{Modality-Specific Pre-Processing and Compression} \label{sc:4}

The UAV's positional information constitutes only a small fraction of the raw sensing data. It is, therefore, desirable to extract and transmit only the relevant features to the BS for processing on the edge server. This helps to optimize the computing and transmission resources between the sensing nodes and the edge server. The data pre-processing at the sensing nodes is specific to the modality they provide. This section describes the data pre-processing steps for each modality considered including the IB-based data compression.

\subsection{Multimodal Data Pre-Processing}

It is useful to first clarify why the multimodal pre-processing methods shown in \fref{fig:dp} are used. These methods aim to reduce the sensing data size while preserving target-related information required for subsequent 3D UAV localization. The existing generic pre-processing methods are not sufficiently adapted for this task, and they suffer from common limitations such as loss of fine-grained 3D geometry, lack of compact target-level representation, and insufficient handling of the background noise and interference \cite{jiang2022lidar, kosuge2022mmwave, 11105406}. Consequently, the task-specific pre-processing has been adopted for different modalities to provide a static background construction and dynamic-static differencing for LiDAR, a point-cloud construction and target clustering for radar, and a fisheye distortion correction with ROI extraction for visual images.

\subsubsection{LiDAR and Radar Modality}

The LiDAR is widely used for detecting objects in the environment. The LiDAR transmits laser pulses, and evaluates their reflections from nearby obstacles. The 3D LiDAR scans both horizontal and vertical directions to produce a 3D point cloud representing the obstacles in the surrounding environment. If an echo pulse reflected in a given spatial direction, $(\theta^\mathrm{L}, \phi^\mathrm{L})$, arrives back with a time delay $\tau^\mathrm{L}$ and the power exceeding a predefined threshold, the distance to the reflector is determined as,
\begin{equation}
  R^{\mathrm{L}}(\theta^\mathrm{L}, \phi^\mathrm{L} ) = \dfrac{1}{2}c \tau^\mathrm{L},
\end{equation}
where, $c=3\times10^8$ \si{m/s}, is the speed of light.

\begin{figure*}[t]
  \centering
  \includegraphics[width=0.7\linewidth, trim=10 10 10 10,
  clip]{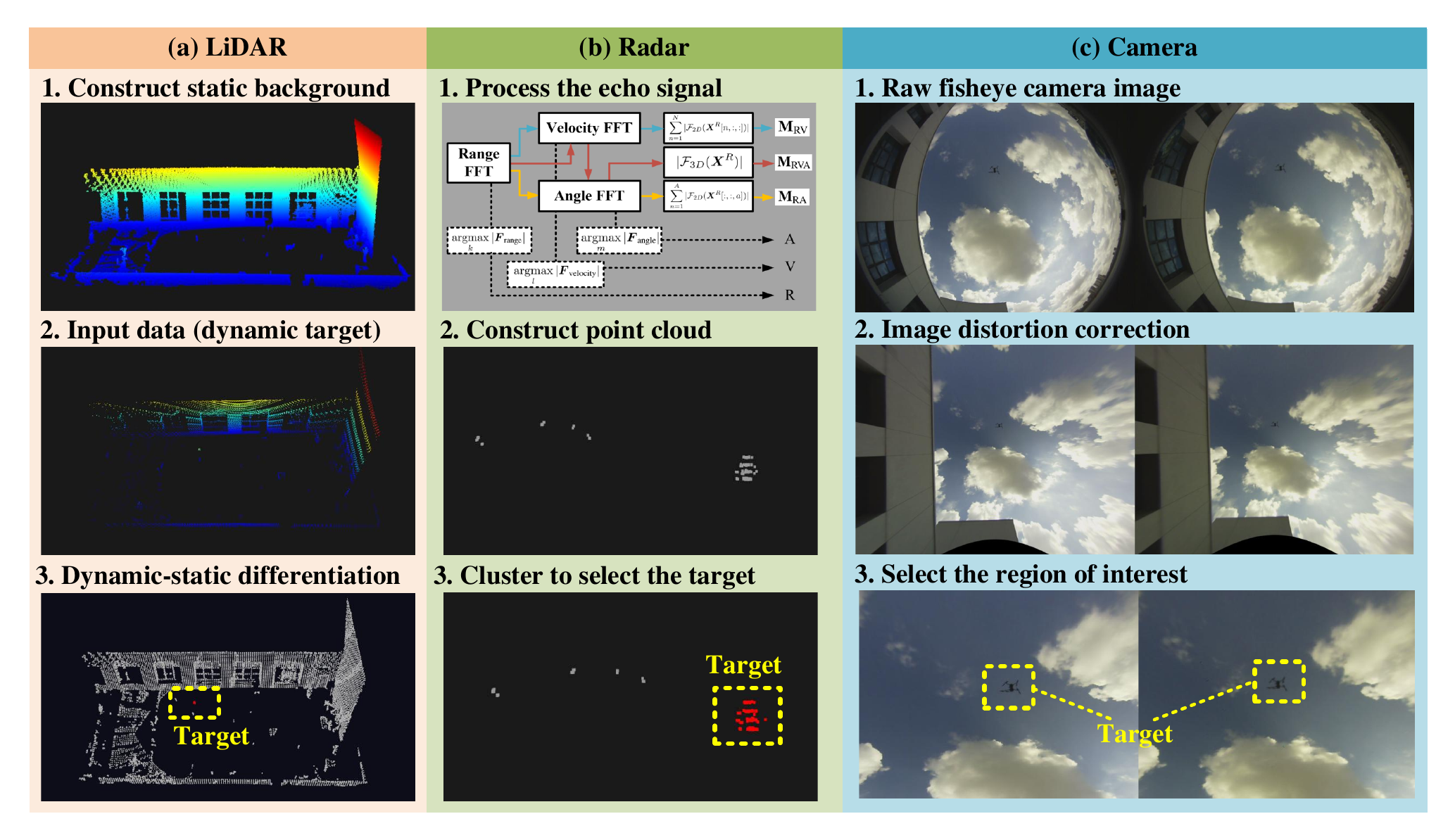}
  \caption{The key steps in the modality-specific data pre-processing for LiDAR, radar, and camera sensors, respectively.}
  \label{fig:dp}
\end{figure*}

In addition to reflections from the target, the LiDAR point cloud contains many reflections from other static objects. The static reflections must be removed in order to isolate the reflections corresponding to the moving target. This can be achieved by analyzing temporal variations in the point cloud. In particular, the point cloud representing the static surroundings can be obtained by performing multiple prior measurements. As indicated in the first column of \fref{fig:dp}, the LiDAR point cloud for the target is simply obtained by subtracting the static background from each LiDAR scan. The newly detected points in the cloud are denoted as, $ \bm{ S}^{\mathrm{L}} \in \mathbb{R}^{ N^{\mathrm{L}} \times 3}$, where $ N^{\mathrm{L}}$ is the number of data points. Each row of $\bm{ S}^{\mathrm{L}}$ contains the polar coordinates, $R^{\mathrm{L}}$, $\theta^\mathrm{L}$, and, $\phi^\mathrm{L}$, of the estimated target location. In some scenarios, the Cartesian coordinates $x^{\mathrm{L}}$, $y^{\mathrm{L}}$, and $z^{\mathrm{L}}$ can be equivalently considered instead of the polar coordinates.

Radar is mainly intended for tracking mobile targets. The key parameters for estimating the target location are obtained by analyzing the sequence of reflected frequency-modulated continuous wave (FMCW) chirps. The FMCW radar directly measures the target range, and the corresponding Doppler shift in different directions. The Doppler component can be used to filter out spurious reflections due to static obstacles. In particular, the instantaneous frequency of each chirp is, $f(t) = f_c + Kt$, where $f_c$ is the center frequency of the carrier, and $K$ is the constant frequency slope. After de-chirping, the beat frequency is linearly proportional to the time delay, $\tau^\mathrm{R}$, i.e., $f_b = K\tau^\mathrm{R}$. The distance to the target can then be computed as,
\begin{equation}
  R^{\mathrm{R}} = \dfrac{c}{2K} f_b.
\end{equation}

The azimuth, $\theta^\mathrm{R}$, and the elevation, $\phi^\mathrm{R}$, of the target can be estimated using a uniform planar array (UPA) with the inter-element spacing, $d_x$, and, $d_y$, respectively. The phase differences between the adjacent UPA elements along the $x$ and $y$ directions can be approximated as,
\begin{equation}
  \begin{split}
    &\Delta \phi_x \approx \dfrac{2\pi d_x}{\lambda_c} \cos{\phi^\mathrm{R}}
    \cos{\theta^\mathrm{R}}, \\
    &\Delta \phi_y \approx \dfrac{2\pi d_y}{\lambda_c} \cos{\phi^\mathrm{R}}
    \sin{\theta^\mathrm{R}},
  \end{split}
\end{equation}
where $\lambda_c=c/f_c$ is the carrier wavelength. The azimuth and the elevation angles can be estimated using the subspace methods, for example, MUSIC and ESPRIT \cite{lavate2010performance}. The spurious dynamic objects, such as birds, can be separated by clustering the points in the radar point cloud. In the considered single-target scenario, the UAV produces the dominant point cluster after the static reflections have been suppressed using Doppler information. Thus, the largest cluster is selected as the target cluster, as indicated in the second column of \fref{fig:dp}. This selection can be performed directly for the radar measurements, and it does not require prior knowledge of the UAV position. The final radar point cloud is denoted as $\bm{X}^{\mathrm{R}}$.

\subsubsection{Vision Modality}\label{ssc:vision}

\begin{figure}[t]
  \centering
  \includegraphics[width=0.95\linewidth]{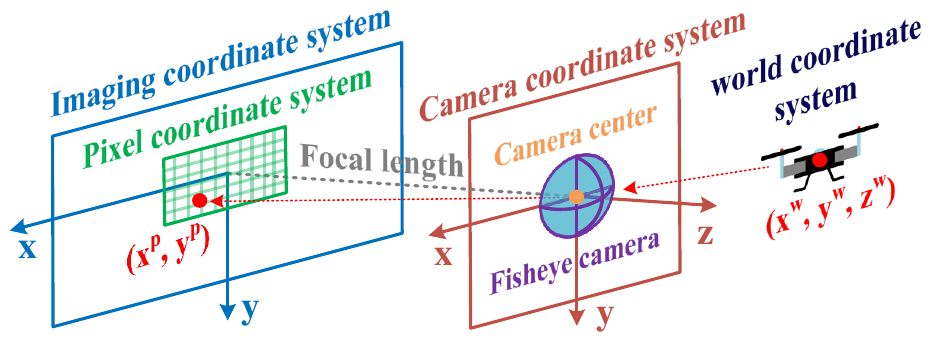} 
  \caption{Derivation of the mapping between the actual UAV coordinates and the image pixel coordinates.}
  \label{fig:camera}
\end{figure}

The likelihood of detecting a UAV can be increased by adding another complementary sensing modality. The fisheye camera offers an ultra-wide field of view at the cost of significant projective distortion. It is also necessary to establish a mapping between the actual UAV coordinates and the image pixels. Such a mapping can be constructed by composing a mapping from the world coordinates to the camera coordinates, a mapping from the camera coordinates to the image coordinates, and a mapping from the image coordinates to the pixel coordinates, as depicted in \fref{fig:camera}.

In particular, a point in the world coordinates, $\bm{p}^{\mathrm{w}}=[x^\mathrm{w}, y^\mathrm{w}, z^\mathrm{w}]^{\top}$, can be converted to the camera coordinates using the rotation, $\bm{R} \in \mathbb{R}^{3 \times 3}$, and the translation, $\bm{t} \in \mathbb{R}^{3 \times 1}$, i.e.,
\begin{equation}
  \bm{p}^{\mathrm{c}} = \bm{R} \bm{p}^{\mathrm{w}} + \bm{t} = [x^\mathrm{c},
  y^\mathrm{c}, z^\mathrm{c}]^{\top}.
\end{equation}
The point, $\bm{p}^{\mathrm{c}}$, is then projected onto the image plane. However, unlike conventional cameras, which employ a linear pinhole projection, the fisheye camera performs a nonlinear projection \cite{mei2007single}. The corresponding point in the image coordinates, $\bm{p}^{\mathrm{i}} = [x^\mathrm{i}, y^\mathrm{i}]^{\top} $, can be expressed as,
\begin{equation}
  \begin{split}
    &x^\mathrm{i} = \dfrac{ f x^\mathrm{c} }{z^\mathrm{c} + \xi r^\mathrm{c}}, \quad
    y^\mathrm{i} = \dfrac{ f y^\mathrm{c}}{z^\mathrm{c} + \xi  r^\mathrm{c}}, \\
    &r^\mathrm{c} = \sqrt{{(x^\mathrm{c})}^2 + {(y^\mathrm{c})}^2+ {(z^\mathrm{c})}^2 },
  \end{split}
\end{equation}
where $\xi$ is the severity of the wide-angle fisheye distortion, $f$ is the focal length, and $r^\mathrm{c}$ is the distance to the camera center. These parameters are normally determined by prior calibration of the camera.

The fisheye camera is also subject to the lens distortion having the radial and tangential components. The image coordinates involving the distortion are expressed as,
\begin{equation} \label{eq:distortions}
  \begin{split}
    x^\mathrm{d} =& x^\mathrm{i}(1+ k_1 {(r^\mathrm{c})}^2 + k_2
    {(r^\mathrm{c})}^4 + k_3 {(r^\mathrm{c})}^6) \\
    &+ 2p_1 x^\mathrm{i} y^\mathrm{i} + p_2({(r^\mathrm{c})}^2+
    {(x^\mathrm{i})}^2), \\
    y^\mathrm{d} =& y^\mathrm{i}(1+ k_1 {(r^\mathrm{c})}^2 + k_2
    {(r^\mathrm{c})}^4 + k_3 {(r^\mathrm{c})}^6) \\
    &+ 2p_1 x^\mathrm{i} y^\mathrm{i} + p_2({(r^\mathrm{c})}^2+
    {(y^\mathrm{i})}^2), 
  \end{split}
\end{equation}
where $k_1$, $k_2$, $k_3$ are the radial distortion coefficients, and $p_1$ and
$p_2$ are the tangential distortion coefficients, respectively.

Finally, let $\alpha$ and $\beta$ denote the number of pixels per millimeter along the $x$ and $y$ axes, respectively, and let $(u,v)$ be the pixel coordinates of the image center. Then, the pixel coordinates, $\bm{p}^{\mathrm{p}} = [x^\mathrm{p}, y^\mathrm{p}]^{\top}$, can be expressed as,
\begin{equation} \label{eq:xp} 
x^\mathrm{p} = \alpha x^\mathrm{d} + u, \quad
y^\mathrm{p} = \beta y^\mathrm{d} + v.
\end{equation}

Obtaining usable images from the fisheye camera requires transforming its non-linear projection to an equivalent linear pinhole projection. The non-linear projection can be corrected by inverting the mapping \eqref{eq:xp} to recover the coordinates, $(x^{\mathrm{d}},y^{\mathrm{d}})$. Removing the distortion \eqref{eq:distortions} requires numerically solving the intrinsic coordinates, $(x^{\mathrm{i}},y^{\mathrm{i}})$, for example, using a method outlined in \cite{santana2016iterative}. The 2D image coordinates can be mapped to the 3D camera coordinates using a nonlinear model described in \cite{mei2007single}. The 3D camera coordinates can be expressed as, $(x^{\mathrm{c}},y^{\mathrm{c}},z^{\mathrm{c}}) = (x^{\mathrm{i}} z^{\mathrm{c}}, y^{\mathrm{i}} z^{\mathrm{c}}, z^{\mathrm{c}})$, where the depth $z$ is computed as,
\begin{align}
  &r^{\mathrm{i}} = \sqrt{{x^{\mathrm{i}}}^2 + {y^{\mathrm{i}}}^2}, \quad 
  z^{\mathrm{c}} = \dfrac{1}{\xi + \sqrt{1 + (1-\xi^2) (r^{\mathrm{i}})^2 }}.
\end{align}
In the final step, the perspective of a linear pinhole camera is projected onto the 2D image using the camera 3D coordinates. The overall fisheye camera perspective rectification process is illustrated in the third column of \fref{fig:dp}. In particular, the UAV footprint occupies only a small portion of the image, so processing the entire image frame is computationally inefficient. At the same time, the sensing devices do not have sufficient computing resources to run complex object detection and tracking algorithms. Hence, we adopt a predefined ROI that can be determined by calibrating the camera in a given deployment environment. Specifically, after the fisheye distortion correction, the invalid image regions that are caused by the rectification process and the persistently occluded regions due to surrounding obstacles are excluded. The remaining effective field-of-view region is used as the fixed ROI in the subsequent feature extraction. We emphasize that such an ROI can be determined offline, and it does not depend on the actual UAV location.

\subsection{IB-based Collaborative Compression}

After the pre-processing, the data volume can be further reduced by removing redundant information that is not relevant to the localization task, for example, by exploiting the IB principle \cite{tishby2000information, hu2024survey}. Removing information redundancy greatly reduces the data transmission requirements. The features of different modalities are compressed in a shared latent space. The shared latent space facilitates subsequent multimodal fusion and target-location estimation.

The IB principle can be described as follows. Given a random input, $X$, with distribution, $p(X)$, and a task-related objective, $Y$, with distribution, $p(Y)$, the IB method seeks a compressed representation $T$ of $X$ that maximally preserves information about $Y$ by discarding redundant information in $X$ that is independent of $Y$. Formally, the IB objective can be formulated in terms of mutual information (MI), $I(X,Y)$, between the random variables, $X$, and, $Y$. The objective is to minimize the information loss,
\begin{equation}
  \mathcal{L}_{\text{IB}} = I(T; X) - \beta I(T; Y),
\end{equation}
where $\beta$ is a hyperparameter, which controls the trade-off between the compression and the task performance. The MI can be computed by the Kullback-Leibler (KL) divergence between the joint distribution, $p(X,Y)$, and the product of the marginal distributions, $p(X)p(Y)$, i.e.,
\cite{mai2022multimodal}
\begin{equation}
  \begin{split}
    I(X; Y) &= D_{\text{KL}}\left( p(x,y) \parallel p(x)p(y) \right) \\
    &= \mathbb{E}_{p(x,y)} \left\{ \log \frac{p(x,y)}{p(x)p(y)} \right\}.
  \end{split}
\end{equation}
In general, calculating $I(T; Y)$ and $I(T; X)$ is difficult, since the underlying distributions are high-dimensional, and often not known.

The case of intractable distributions is often addressed by adopting variational methods. The variational information bottleneck (VIB) finds the optimum approximations of the underlying distributions by iteratively improving the variational bounds. It also employs a re-parameterization trick to allow differentiable sampling, which makes the end-to-end neural network training feasible \cite{alemi2016deep}. In particular, the VIB assumes a learnable stochastic encoder, $q_{\phi}(t|x) \sim \mathcal{N}(\mu(x), \sigma^2(x))$, and a decoder, $p_{\theta}(y|t)$, having the Gaussian prior, $p(t) \sim \mathcal{N}(\bm{0}, \bm{I})$. The intractable terms in $I(T; Y)$ and $I(T; X)$, respectively, are then replaced by the variational lower and upper bounds, i.e.,
\begin{equation} \label{eq:I_T_X}
  \begin{split}
    I(T; Y) &\geq \mathbb{E}_{p(x,y)} \mathbb{E}_{ q_{\phi}(t|x)} \left[ \log
      p_{\theta}(y|t) \right], \\
    I(T; X) &\leq \mathbb{E}_{p(x)} \left[ D_{\text{KL}} ( q_{\phi}(t|x) ||
      p(t) ) \right].
  \end{split}
\end{equation}
The re-parameterization trick, $ t = \mu_{\phi}(x) + \sigma_{\phi}(x) \odot \epsilon , \epsilon \sim \mathcal{N}(0,I)$, is used to enable a low-variance, differentiable sampling from the Gaussian encoder in order to make the VIB process feasible.

\begin{figure*}[t]
  \centering
  \includegraphics[width=0.85\linewidth, trim=10 15 10 10,
  clip]{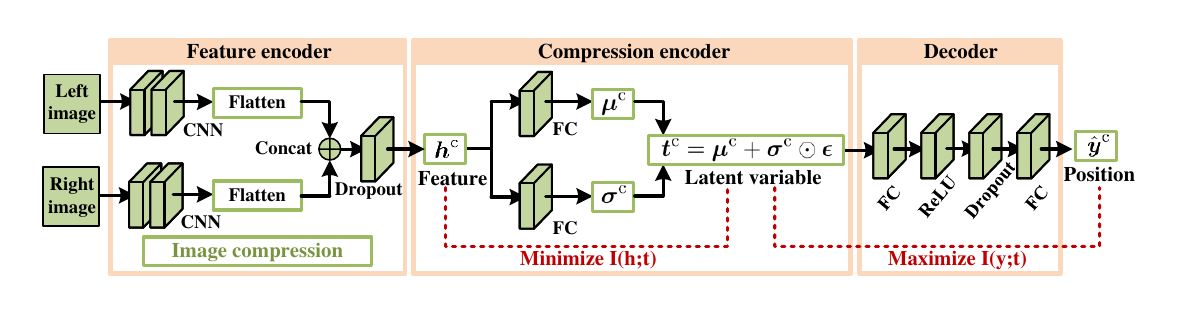}
  \caption{The architecture of the proposed VIB-based compression module.}
  \label{fig:cn}
\end{figure*}

The architecture of the proposed VIB-based image compression module is shown in \fref{fig:cn}. The module employs two encoders: a feature encoder and a compression encoder. The decoder is used only during training in order to facilitate the model learning. The feature encoder extracts visual representations from the input image frames; a CNN has been chosen for this encoder. In the case of stereoscopic camera, the left and the right images are processed by two parallel CNN branches to generate the feature vectors, $\bm{h}^{\mathrm{left}}$, and, $\bm{h}^{\mathrm{right}}$. These features are then concatenated and passed through a dropout layer to obtain the final visual representation,
\begin{equation}
  \bm{h}^{\mathrm{c}} = \mathrm{Dropout}\left(
    \mathrm{Concat}(\bm{h}^\mathrm{left}, \bm{h}^\mathrm{right}) \right).
\end{equation}

The visual representation, $\bm{h}^{\mathrm{c}}$, is used to generate a compressed latent variable, $\bm{t}^{\mathrm{c}}$. Specifically, the mean, $\bm{\mu}^{\mathrm{c}}$, and the standard deviation, $\bm{\sigma}^{\mathrm{c}}$, are generated by two multi-layer perceptrons (MLPs), so the latent representation becomes,
\begin{equation}
  \bm{t}^{\mathrm{c}}  = \bm{\mu}^{\mathrm{c}} + \bm{\sigma}^{\mathrm{c}} \odot
  \bm{\epsilon} \in \mathbb{R}^{z^\mathrm{c}},
\end{equation}
where $\bm{\epsilon} \sim \mathcal{N}(\bm{0}, \bm{I})$ is a sample from the standard Gaussian distribution. Since $q_{\phi}(\bm{t}^{\mathrm{c}} |\bm{h}^{\mathrm{c}})$ is modeled as a diagonal Gaussian distribution and $p(\bm{t}^{\mathrm{c}})$ is adopted as the prior, the KL-divergence term can be computed in closed form as,
\begin{equation}
  D^{\mathrm{c}}_{\mathrm{KL}}\!\big(q_\phi(\bm{t}^{\mathrm{c}}
  |\bm{h}^\mathrm{c})\Vert p(\bm{t}^{\mathrm{c}}))\big)
  = \frac{1}{2} \sum_{j=1}^{z^{\mathrm{c}}} \big(
  {\bm{\mu}^{\mathrm{c}}_{j}}^2 + {\bm{\sigma}^{\mathrm{c}}_{j}}^2 - \log
  {\bm{\sigma}^{\mathrm{c}}_{j}}^2 - 1 \big).
\end{equation}

After constructing the two encoders, a decoder is introduced during training to ensure that the encoded representations preserve task-relevant information. Assuming UAV localization as the main task, the decoder is designed to output the estimated 3D location, $\hat{\bm{y}}^{\mathrm{c}} \in \mathbb{R}^3$. The decoder is again implemented as a MLP. The decoder likelihood is modeled as,
\begin{equation}
  p_\theta(\bm{y}| \bm{t}^{\mathrm{c}})=
  \mathcal{N}(\hat{\bm{y}}^{\mathrm{c}}, \sigma_y^2\bm{I}),
\end{equation}
where $\sigma_y^2$ is the variance of the observation noise. Consequently,
\begin{equation}
  -\log p_\theta(\bm{y}| \bm{t}^{\mathrm{c}}) \propto \lVert \bm{y} -
  \hat{\bm{y}}^{\mathrm{c}} \rVert_2^2 = \mathcal{L}^c_\mathrm{MSE},
\end{equation}
and the overall loss function is expressed as,
\begin{equation}
  \mathcal{L}^{\mathrm{c}}_\mathrm{ALL} 
  = \underbrace{\mathcal{L}^{\mathrm{c}}_\mathrm{MSE}}_{\text{task loss}} 
  + \beta \underbrace{
    D^{\mathrm{c}}_{\mathrm{KL}}\!\big(q_\phi(\bm{t}^{\mathrm{c}}
    |\bm{h}^\mathrm{c})\Vert p(\bm{t}^{\mathrm{c}}))\big)  }_{\text{information compression}}.
\end{equation}
This loss function optimizes the IB criterion, which encourages the visual features, $\bm{t}^{\mathrm{c}}$, to preserve information that is relevant to the UAV location while other redundant details are discarded.

The described IB-based compression framework can be directly extended to the LiDAR and radar modalities. Since both these modalities produce point clouds, the PointNet architecture \cite{qi2017pointnet} has been adopted as the feature encoder. The compression encoders, decoder, and the loss function remain the same as for the vision modality. The detailed configuration and the parameter settings of the compression network are provided in the next section.

\section{Multimodal Data Alignment and Fusion} \label{sc:5}

\begin{figure*}[t]
  \centering
  \includegraphics[width=0.8\linewidth, trim=10 15 10 10,
  clip]{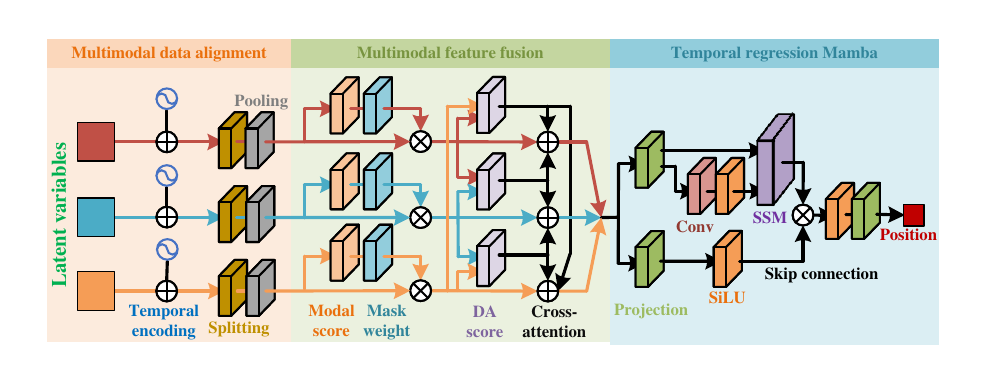}
  \caption{The data processing network for collaborative UAV localization using
    multiple modal sensing sources.}
  \label{fig:fusion}
\end{figure*}

The UAV localization is inferred from the time-aligned and fused data obtained from different modal sensing sources. Both these tasks are performed by two separate modules. The fusion module is also used to handle low-quality and missing data. The overall data processing network for UAV localization is shown in \fref{fig:fusion}. It leverages the temporal correlations in order to enhance the localization accuracy. The network consists of three main modules: multimodal data alignment, multimodal feature fusion, and temporal regression; these modules are described next.

\subsection{Time-Alignment Module}

The edge server receives the modal features representing different sensing modalities. Recall that these features are compressed representations of the raw modal data, which were obtained at the sensing nodes using the IB-based compression as discussed above. Assuming camera, LiDAR, and radar as modal sensing sources, their low-dimensional latent representations are, respectively, denoted as,
\begin{equation}
  \bm{t}^{\mathrm{c}}_i \in \mathbb{R}^{z^{\mathrm{c}}}, \quad
  \bm{t}^{\mathrm{l}}_i \in \mathbb{R}^{z^{\mathrm{l}}}, \quad
  \bm{t}^{\mathrm{r}}_i \in \mathbb{R}^{z^{\mathrm{r}}}.
\end{equation}
For simplicity, the superscript, $\mathrm{m} \in \{{\mathrm{c}},{\mathrm{l}},{\mathrm{r}}\}$, will be used to denote the specific modality type, i.e., camera, LiDAR, or radar, respectively. More importantly, different modalities have different sampling rates denoted as, $f^{\mathrm{c}}$, $f^{\mathrm{l}}$, and $f^{\mathrm{r}}$. The disparity in sampling rates causes temporal misalignment, uneven information density, noise amplification during fusion, and subsequently, the model training instability and reduced localization accuracy. These effects can be mitigated by aligning the data streams from different modal sources to a common time reference and a common sampling rate. This task is performed by the designed TE-Align module, which produces a timestamp-guided window-level temporal representation alignment of the incoming heterogeneous sensing streams.

Let $\tau_i^{\mathrm{m}}$ denote a timestamp of the $i$-th data frame for modality, $m$. One strategy for aligning different modalities is to adopt a modality-specific encoding that injects temporal information into its feature representation. Such an encoding can be defined as,
\begin{equation}
  \begin{split}
    \bm{E}^{\mathrm{m}}_i &= \left[
      \sin\left(\dfrac{\tau_i^{\mathrm{m}}}{\kappa^{2j/z^{\mathrm{m}}}}\right),
      \cos\left(\dfrac{\tau_i^{\mathrm{m}}}{\kappa^{2j/z^{\mathrm{m}}}}\right)
    \right]_{j=1}^{z^{\mathrm{m}}/2}, \\
    \bm{t}^{\mathrm{m},\mathrm{e}}_i  &=  \bm{t}^{\mathrm{m}}_i +
    \bm{E}^{\mathrm{m}}_i \in \mathbb{R}^{z^{\mathrm{m}}}.
  \end{split}
\end{equation}

The modalities with different sampling rates are mapped to a common time grid, and then segmented into the fixed windows of length, $t^{\mathrm{w}}$, so they can be concatenated for the subsequent downstream processing. The sample pooling within windows suppresses the noise even when some modal frames are dropped or are missing, which also significantly reduces the computational and memory overheads. In particular, let time indices of the $k$-th window be the subsets, $\bm{s}^{\mathrm{m}}(k) = \{ i : \tau_i^{\mathrm{m}} \in \bm{n}_k \}$, where $ \bm{n}_k = \{(k-1)t^{\mathrm{w}}, kt^{\mathrm{w}}\}$. The pooling over the $k$-th time window with time-alignment can be expressed as,
\begin{equation}
  \bm{t}^{\mathrm{m},\mathrm{e}}(k) = \dfrac{1}{|\bm{s}^{\mathrm{m}}(k)|}
  \sum_{i \in \bm{s}^{\mathrm{m}}(k)} \bm{t}^{\mathrm{m},\mathrm{e}}_i, \quad
  |\bm{s}^{\mathrm{m}}(k)|>0. 
\end{equation}
It should be emphasized that the resulting representation, $\bm{t}^{\mathrm{m},\mathrm{e}}(k)$, is a window-level temporal representation rather than a frame-level synchronized observation. Note also that, when there is no observation for modality, $m$, within the $k$-th temporal window, then $|\bm{s}^{\mathrm{m}}(k)|=0$, and the corresponding latent representation and the availability mask are set to zero. The unavailable modality can then be excluded from the fusion in the MADW module.

\subsection{Multimodal Fusion Module}

After the time-alignment, the data from different modal sources need to be fused together. The data fusion is accomplished using two modules, i.e., MADW and DACA.

\subsubsection{Mask-Aware Dynamic Weighting (MADW)}

The MADW module addresses the problem of degraded and occasionally missing modalities. However, instead of explicitly estimating the reliability of physical sensing, it learns the task-dependent modality confidence scores from the aligned latent representations, which indicate the relative contributions of each modality to the main localization task. The confidence scores, $s^{\mathrm{m}}(k) = \text{score}^{\mathrm{m}}(\bm{t}^{\mathrm{m},\mathrm{e}}(k))$, are obtained using a MLP-based computation. Note that these scores cannot be obtained using supervised learning, since no explicit sensing-quality labels are available. The scoring network is optimized jointly with the localization objective in order to learn which modal representations are the most beneficial for UAV localization under different sensing conditions.

In order to support handling of missing data, the data availability mask is introduced. In particular, $\text{mask}^{\mathrm{m}}(k)=1$, if the modality is available, and $\text{mask}^{\mathrm{m}}(k)=0$, otherwise. The overall reliability weights are defined assuming the masked softmax function, i.e.,
\begin{equation}
  \begin{split}
    w^{\mathrm{m}}(k) =&\  \text{softmax}\big(\text{mask}^{\mathrm{m}}(k) \cdot
    s^{\mathrm{m}}(k) \\ &+ (1 - \text{mask}^{\mathrm{m}}(k)) \cdot
    (-\infty)\big).
  \end{split}
\end{equation}
The softmax function acts as a simple and interpretable gating mechanism. It dynamically adjusts the contributions of different modalities, while suppressing less informative or unavailable modal representations. The contributions of different modalities are differentiated by scaling the modal features as,
\begin{equation}
  \bm{t}^{\mathrm{m},\mathrm{w}}(k) = w^{\mathrm{m}}(k) \cdot
  \bm{t}^{\mathrm{m},\mathrm{e}}(k).
\end{equation}

\subsubsection{Dependency-Aware Cross-Attention}

In conventional transformers, the cross-attention derives the attention weights from dot-product similarity. It reflects a linear geometric similarity between the projected features. However, for noisy, imbalanced, misaligned, or partially missing modality conditions, the dot-product similarity scores become disturbed by the noise-induced cross terms, so they fail to capture the task-relevant dependencies. The DACA addresses this problem by replacing a dot-product similarity with a learnable dependency score. It enables nonlinear modeling of the task-relevant cross-modal dependencies, thereby improving the robustness of the subsequent multimodal fusion.

The IB-CC retains task-relevant latent information, while suppressing modality-specific redundant information. The compressed latent representation provides a suitable basis for modeling the cross-modal dependencies. The DACA uses the learnable dependency scores to characterize the task-relevant statistical associations between the compressed modal representations. In particular, assuming the weighted latent representations, $\bm{a}$, and, $\bm{b}$, the dependency score is defined as,
\begin{equation}
    g_{\eta}(\bm{a},\bm{b}) = \bm{w}_{2}^{\top} \sigma \left( \bm{W}_{1} \left[
        \bm{a}, \bm{b}, \bm{a}-\bm{b}, \bm{a}\odot\bm{b}
      \right] + \bm{b}_{1} \right) + b_{2},
\end{equation}
where $\eta=\{\bm{W}_{1},\bm{w}_{2},\bm{b}_{1},b_{2}\}$ denotes the set of learnable parameters of the MLP scorer, $\sigma(\cdot)$ is a nonlinear activation function, $[\cdot]$ denotes concatenation, and $\odot$ is the element-wise multiplication. The terms, $\bm{a}-\bm{b}$, and, $\bm{a}\odot\bm{b}$, are introduced to describe the discrepancy and the interaction between the modalities, respectively.

Assuming camera as the target modality, the DACA computes the pairwise dependency scores with LiDAR and radar, respectively, i.e.,
\begin{align}
  s^{\mathrm{c}\leftarrow\mathrm{l}}(k) = g_{\eta}
  \left( \bm{t}^{\mathrm{c},\mathrm{w}}(k), \bm{t}^{\mathrm{l},\mathrm{w}}(k)
  \right), \\
  s^{\mathrm{c}\leftarrow\mathrm{r}}(k) = g_{\eta} \left(
    \bm{t}^{\mathrm{c},\mathrm{w}}(k), \bm{t}^{\mathrm{r},\mathrm{w}}(k)
  \right).
\end{align}
The corresponding cross-modal attention weights are obtained using the softmax, i.e.,
\begin{align}
  \alpha^{\mathrm{c}\leftarrow\mathrm{l}}(k) = \frac{ \exp
    \left(s^{\mathrm{c}\leftarrow\mathrm{l}}(k)\right) }{ \exp
    \left(s^{\mathrm{c}\leftarrow\mathrm{l}}(k)\right) + \exp
    \left(s^{\mathrm{c}\leftarrow\mathrm{r}}(k)\right) }, \\
  \alpha^{\mathrm{c}\leftarrow\mathrm{r}}(k) = \frac{\exp
    \left(s^{\mathrm{c}\leftarrow\mathrm{r}}(k)\right) }{ \exp
    \left(s^{\mathrm{c}\leftarrow\mathrm{l}}(k)\right) + \exp
    \left(s^{\mathrm{c}\leftarrow\mathrm{r}}(k)\right) }.
\end{align}

The refined camera features are then computed by injecting dependency-weighted information from the other two modalities, i.e.,
\begin{equation}
  \begin{split}
    \bm{t}^{\mathrm{c},\mathrm{f}}(k) =&\ \bm{t}^{\mathrm{c},\mathrm{w}}(k) +
    \lambda^{\mathrm{c},\mathrm{l}}
    \alpha^{\mathrm{c}\leftarrow\mathrm{l}}(k) \bm{t}^{\mathrm{l},\mathrm{w}}(k) \\
    &+ \lambda^{\mathrm{c},\mathrm{r}}
    \alpha^{\mathrm{c}\leftarrow\mathrm{r}}(k)
    \bm{t}^{\mathrm{r},\mathrm{w}}(k),
  \end{split}
\end{equation}
where, $\lambda^{\mathrm{c},\mathrm{l}}$, and, $\lambda^{\mathrm{c},\mathrm{r}}$, are the learnable scaling factors.

The refined features for LiDAR and radar are obtained analogously. Finally, the three refined modality features are concatenated, and their fused representation is expressed as,
\begin{equation}
  \bm{t}^{\mathrm{f}}(k) = [ \bm{t}^{\mathrm{c},\mathrm{f}}(k),
  \bm{t}^{\mathrm{l},\mathrm{f}}(k), \bm{t}^{\mathrm{r},\mathrm{f}}(k) ].
\end{equation}

\subsection{Mamba-based Localization Module}

The Mamba architecture has been adopted for modeling the temporal dependencies in the fused multimodal representation, $\bm{t}^{\mathrm{f}}(k)$. The goal is to provide continuous target tracking. The Mamba architecture approximates the recurrent and convolutional neural network dynamics using a linear state-space model \cite{gu2024mamba}. Thus, given a sequence of the fused multimodal features, $\{\bm{t}^{\mathrm{f}}(1),\bm{t}^{\mathrm{f}}(2),...,\bm{t}^{\mathrm{f}}(K) \}$, the Mamba model maintains the latent state, $\bm{h}(k)$, which evolves as,
\begin{equation}
  \begin{split}
    &\bm{h}(k) = \bm{A}(k) \bm{h}(k-1) + \bm{B}(k) \bm{t}^{\mathrm{f}}(k), \\
    &\bm{y}(k) = \bm{C}(k) \bm{h}(k) + \bm{D}(k) \bm{t}^{\mathrm{f}}(k), \\
    &\{\bm{A}(k),  \bm{B}(k),  \bm{C}(k),  \bm{D}(k)\} =
    g_{\theta}(\bm{t}^{\mathrm{f}}(k)),
  \end{split}
\end{equation}
where $\bm{A}(k) $, $\bm{B}(k) $, $\bm{C}(k) $, and $\bm{D}(k)$ are the input-dependent transition matrices. The gating function, $g_{\theta}(\cdot)$, is implemented using linear projections. This allows the model to adaptively emphasize informative patterns, and suppress information patterns that are less relevant.

The target location tracking represents a regression problem. It minimizes the mean-square error (MSE) between the predicted location, $\bm{\hat{p}}(k)$, and the ground truth, $\bm{p}(k)$, i.e.,
\begin{equation}
  \mathcal{L}^\mathrm{loc} = \dfrac{1}{K} \sum_{k=1}^K  ||\bm{p}(k) -
  \bm{\hat{p}}(k) ||_{2}^{2}.
\end{equation}
The MSE regression is sufficient to capture the long-term temporal correlations as well as the smooth motion trajectories.

\section{Numerical Results} \label{sc:6}

The proposed collaborative UAV localization framework has been evaluated numerically. The goal is to assess the performance of the proposed framework under diverse conditions with different dynamics, partial occlusions, and missing modalities. The experiments utilize the real-world multimodal MMAUD dataset \cite{yuan2024mmaud} that provides rich and high-fidelity sensing data including stereo vision, multi-LiDAR, radar, and the Leica-referenced measurements. 70\% of the data are used for model training, and the remaining data are split equally for model validation and testing. All experiments were conducted on a standard PC equipped with Intel i7-12700H CPU, and NVIDIA RTX 3070 Ti GPU. The training parameters are set as $f^{\mathrm{c}} = 30\,$ \si{Hz}, $f^{\mathrm{l}} = 10\,$ \si{Hz}, $f^{\mathrm{r}} = 15\,$ \si{Hz}, $\beta = 10^{-3}$, and $\lambda = 0.5$. The alignment window length is $0.2$ \si{s}, the learning rate is $10^{-4}$, and the dropout rate is $0.1$. The configurations of different modules are summarized in \tref{ns}.

\begin{table}[t]
  \caption{Module configurations.}
  \label{ns}
  \resizebox{\linewidth}{!}{
    \begin{tabular}{|c|c|}
      \hline \textbf{Module} & \textbf{Construction} \\ \hline
\begin{tabular}[c]{@{}c@{}}Image feature \\ encoder\end{tabular} &
\begin{tabular}[c]{@{}c@{}}$\mathrm{ReLU}\left( \mathrm{Conv}_{3\times
      3,\,s=2,\,p=1}^{(1 \rightarrow 32)}(\cdot) \right) \rightarrow$
  $\mathrm{ReLU}\left( \mathrm{Conv}_{3\times 3,\,s=2,\,p=1}^{(32 \rightarrow
      64)}(\cdot) \right)$.\end{tabular} \\ \hline
\begin{tabular}[c]{@{}c@{}}Point cloud \\ feature encoder\end{tabular} &
\begin{tabular}[c]{@{}c@{}}$\mathrm{ReLU}\!\left( \mathrm{BN}\left(
      \mathrm{Conv1d}_{P_{l-1} \rightarrow P_l}(\mathbf{H}_{l-1}) \right)
  \right)$, \, l = 1, \dots, 5,\\
  $\mathrm{MaxPool}_{ \mathrm{global}} (\mathbf{H}_5) \in
  \mathbb{R}^{1024}$.\end{tabular} \\ \hline IB decoder &
$\mathrm{MLP}_{64} (\mathrm{Dropout} ( \mathrm{ReLU}( \mathrm{MLP}_{128}
(\cdot))) )$. \\ \hline Data align &
$\mathrm{AvgPool}_{t^{\mathrm{w}}} ((\cdot) + \bm{E}^{\mathrm{m}}_i) $. \\
\hline
\begin{tabular}[c]{@{}c@{}}Mask-aware \\ dynamic weighting\end{tabular} &
$\text{softmax} [\text{mask}^{\mathrm{m}}(k) \mathrm{MLP}_{128}(\cdot)+ (1 -
\text{mask}^{\mathrm{m}}(k)) (-\infty) ] (\cdot)$. \\ \hline
\begin{tabular}[c]{@{}c@{}}Dependency \\ cross-attention\end{tabular} &
$(\cdot)^{\mathrm{m},\mathrm{w}} + \sum_{n \neq m} \mathrm{softmax}[
\mathrm{MLP}_{128}((\cdot)^{\mathrm{m},\mathrm{w}},(\cdot)^{\mathrm{n},\mathrm{w}})](\cdot)^{\mathrm{n},\mathrm{w}}$.
\\ \hline
\end{tabular}}
\end{table}

The results are summarized in the following three subsections. The first subsection evaluates the performance of data pre-processing and compression. The second subsection investigates the performance of the complete proposed architecture for UAV localization. Finally, the last subsection considers a communication scenario in which the UAV acts as a communication relay.

\begin{figure}[t]
  \centering
  \begin{subfigure}{0.7\linewidth}
    \centering
    \includegraphics[width=\linewidth]{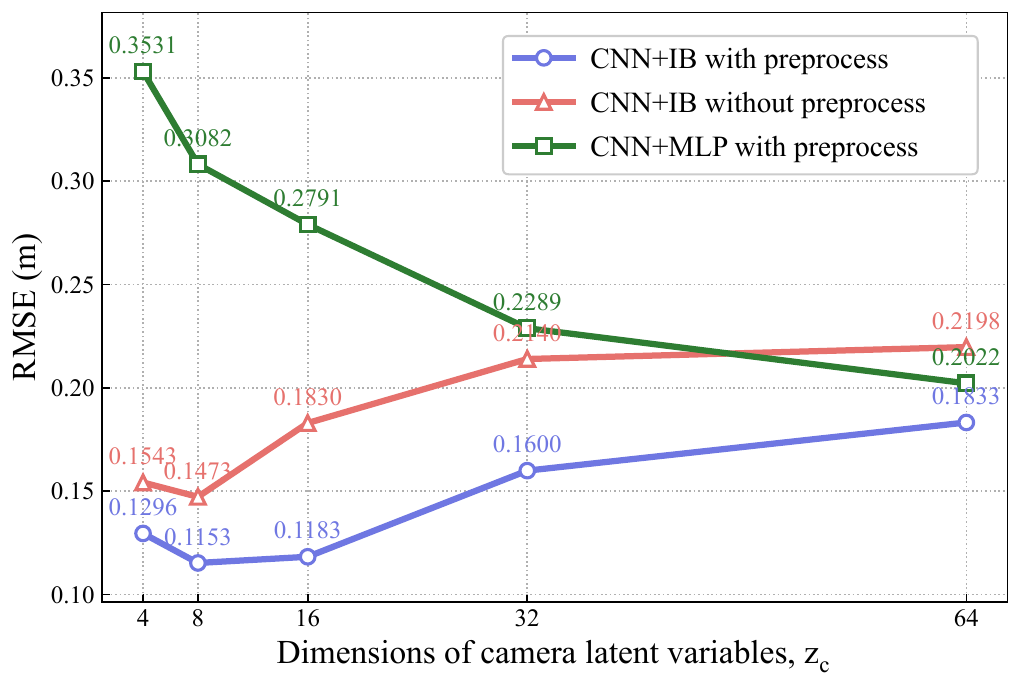}
    \captionsetup{skip=-1pt}
    \caption{Localization errors using only camera modality.}
    \label{zcam}
  \end{subfigure}
  \medskip
  
  \begin{subfigure}{0.7\linewidth}
    \centering
    \includegraphics[width=\linewidth]{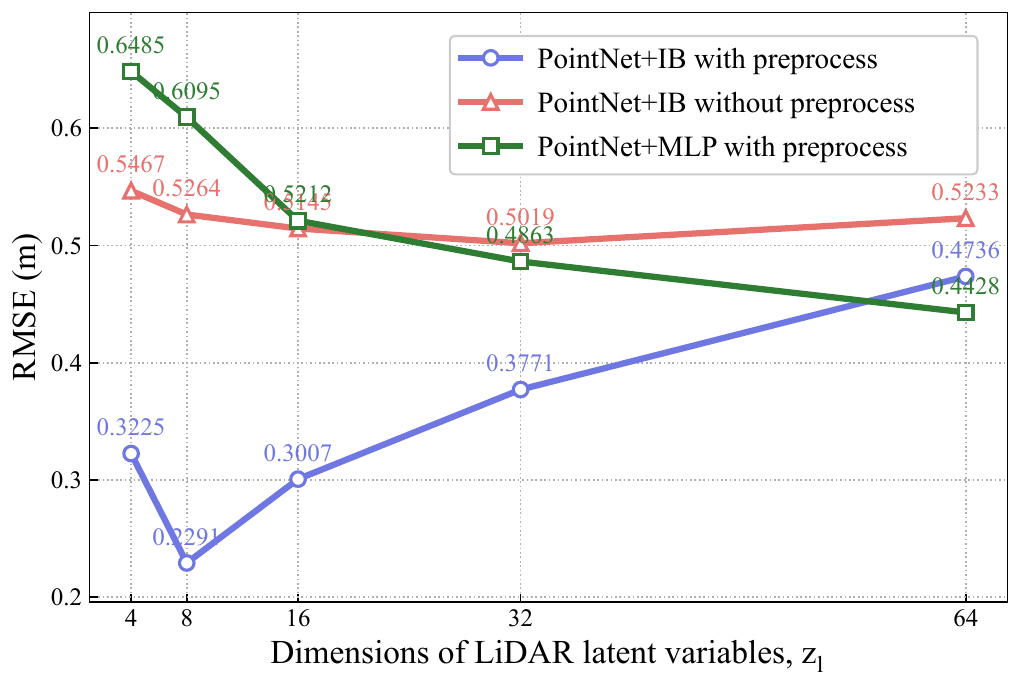}
    \captionsetup{skip=-1pt}
    \caption{Localization errors using only LiDAR modality.}
    \label{zlidar}
  \end{subfigure}
  \medskip
  
  \begin{subfigure}{0.7\linewidth}
    \centering
    \includegraphics[width=\linewidth]{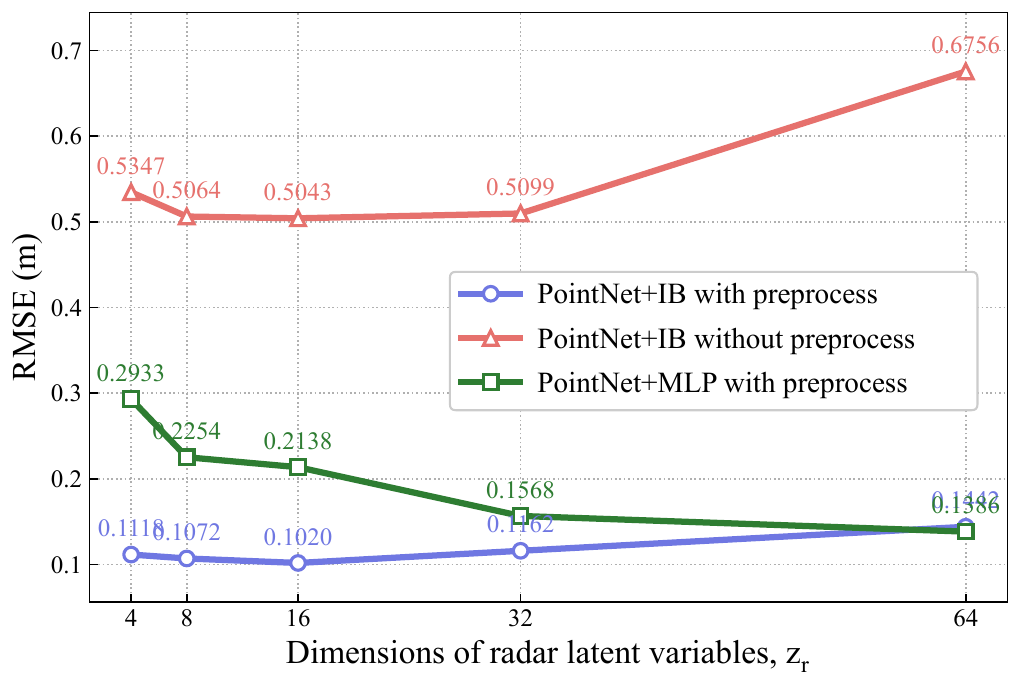} 
    \captionsetup{skip=-1pt}
    \caption{Localization errors using only radar modality.}
    \label{zradar}
  \end{subfigure}
  \caption{The localization errors under different modalities and compression
    methods.}
  \label{z3}
\end{figure}

\begin{table*}[t]
  \centering
  \caption{Comparison of resulting data rates of different modalities before
    and after the data pre-processing and compression.}
  \label{data}
  \resizebox{0.85\linewidth}{!}{
    \begin{tabular}{|c|c|c|ccccc|}
      \hline \multirow{2}{*}{\textbf{Type}} &
      \multirow{2}{*}{\textbf{Original} (KB/s)} &
      \multirow{2}{*}{\textbf{Pre-processed} (KB/s)} &
      \multicolumn{5}{c|}{\textbf{IB-based compression} (KB/s)} \\ \cline{4-8}
      &&&
      \multicolumn{1}{c|}{$z=4$} & \multicolumn{1}{c|}{$z=8$} &
      \multicolumn{1}{c|}{$z=16$} & \multicolumn{1}{c|}{$z=32$} & $z=64$
      \\ \hline
      Camera & $63,701$  (100\%) & $13,261$ (20.82\%) &
      \multicolumn{1}{c|}{$0.94$ (0.15\textpertenthousand)} &
      \multicolumn{1}{c|}{$1.88$ (0.29\textpertenthousand)} &
      \multicolumn{1}{c|}{$3.75$ (0.59\textpertenthousand)} &
      \multicolumn{1}{c|}{$7.50$ (1.18\textpertenthousand)} & $15.00$
      (2.35\textpertenthousand) \\ \hline LiDAR & 4685 (100\%) & $1.87$
      (0.39\%) & \multicolumn{1}{c|}{$0.31$ (0.07\textperthousand)} &
      \multicolumn{1}{c|}{$0.63$ (0.13\textperthousand)} &
      \multicolumn{1}{c|}{$1.25$ (0.27\textperthousand)} &
      \multicolumn{1}{c|}{$2.50$ (0.53\textperthousand)} & $5.00$
      (1.07\textperthousand) \\ \hline Radar & $111$ (100\%) & $31.74$
      (28.57\%) & \multicolumn{1}{c|}{$0.47$ (0.42\%)} &
      \multicolumn{1}{c|}{$0.94$ (0.84\%)} & \multicolumn{1}{c|}{$1.88$
        (1.69\%)} & \multicolumn{1}{c|}{3.75 (3.38\%)} & 7.50 (6.75\%) \\
      \hline 
    \end{tabular} }
\end{table*}

\subsection{Performance of Data Pre-Processing and Compression}

The goal is to evaluate the performance of the proposed modality-specific pre-processing modules, and also the effectiveness of the IB-based compression. The architecture of the proposed compression module has been shown in \fref{fig:cn}. Note that the fusion module is not considered in these investigations. For comparison, the following two baseline schemes are considered: (i) a scheme where raw sensing data are neither pre-processed nor subjected to explicit feature extraction, and, (ii) a scheme where the devised IB module is replaced with a general MLP having the same depth and the same total number of parameters.

Assuming first the camera modality, \fref{zcam} reports the localization errors as the root MSE (RMSE) values across varying latent dimensions when the pre-processing is either included or excluded. It can be observed that the proposed IB-based compression is effective across all settings. However, the performance degrades substantially when the pre-processing is excluded. It can be also observed that the IB-based compression method is superior to the general MLP method. Note also that the IB compression attains a flat minimum in RMSE, since small latent spaces discard task-relevant information, whereas large latent spaces preserve unnecessary redundancy. If the IB compression is replaced with MLP, the localization errors decrease with the size of the latent dimensionality; this suggests that MLP is effective for compressing high-dimensional latent spaces. However, high-dimensional latent spaces also incur large computational cost and transmission overhead, which is rather problematic for the resource-constrained sensing devices.

The RMSE localization errors for LiDAR and radar are shown in \fref{zlidar} and \fref{zradar}, respectively, as a function of the latent space dimension. The localization scheme without a fusion module is again considered. From both these figures, it is evident that the proposed IB-based compression achieves the best or nearly the best performance. Moreover, the scheme comprising the pre-processing consistently outperforms the scheme without pre-processing. It highlights the importance of including the proposed pre-processing. For camera modality, the IB-based compression outperforms the MLP at low-dimensional latent spaces, and the former again experiences a flat minimum.

\begin{table}[t]
  \centering
  \caption{The estimated computational complexities (FLOPs) of different
    modules.}
  \label{flops}
  \resizebox{0.8\linewidth}{!}{
    \begin{tabular}{|c|ccc|cc|}
      \hline \multirow{2}{*}{Module} & \multicolumn{3}{c|}{Pre-processing} &
      \multicolumn{2}{c|}{Encoder} \\ \cline{2-6} & \multicolumn{1}{c|}{Camera}
      & \multicolumn{1}{c|}{LiDAR} & Radar & \multicolumn{1}{c|}{Image} & Point
      cloud \\ \hline MFLOPs & \multicolumn{1}{c|}{207.09} &
      \multicolumn{1}{c|}{14.81} & 40.33 & \multicolumn{1}{c|}{147.32} &
      1426.98 \\ \hline
    \end{tabular} }
\end{table}

The data compression ratios achieved by the proposed IB-based compression method are reported in \tref{data}. The data volumes are measured in kilobytes per second (KB/s). In particular, as shown in \tref{data}, the raw data throughput of camera, LiDAR, and radar modalities are of the order of tens of thousands, thousands, and hundreds of KB/s, respectively. The pre-processing reduces the camera and the radar data rates to 20-30\% of their original volume. The pre-processing of LiDAR data attains much larger reductions as the majority of the data points represent static background, and they are removed from the cloud. The subsequent IB-based compression offers further reduction in the data rates for all the modalities considered down to only a few KB/s.

It is also useful to evaluate whether the proposed methods can be deployed on low-power devices such as Raspberry Pi 5, which provides a computational power of approximately $150$-$200$ GFLOPs \cite{fezari2023raspberry}. Assuming the input image of $128\times128$ pixels, and up to $1024$ points in the point cloud, the estimated computational complexities of the encoder modules and of the developed pre-processing methods are summarized in \tref{flops}.

The theoretical latency for both the image and the point cloud processing pipelines is below $0.01$ \si{s}. Since the actual latency also depends on the hardware utilization, memory access, and software implementation, we further evaluate the developed processing pipelines using actual implementation on a Raspberry Pi 5. The measured latencies of the camera and the point cloud pipelines are approximately $0.02$ \si{s} and $0.09$ \si{s}, respectively. Although these values are larger than the ideal estimates in \tref{flops}, they remain within the adopted alignment-window duration of $0.2$ \si{s}, indicating that the proposed sensing-side processing pipelines are feasible for the considered deployment scenarios. Moreover, these latencies can be further reduced using software optimization, parallel implementation, and hardware-specific acceleration.

\begin{figure}[t]
  \centering
  \begin{subfigure}{0.7\linewidth}
    \centering
    \includegraphics[width=\linewidth]{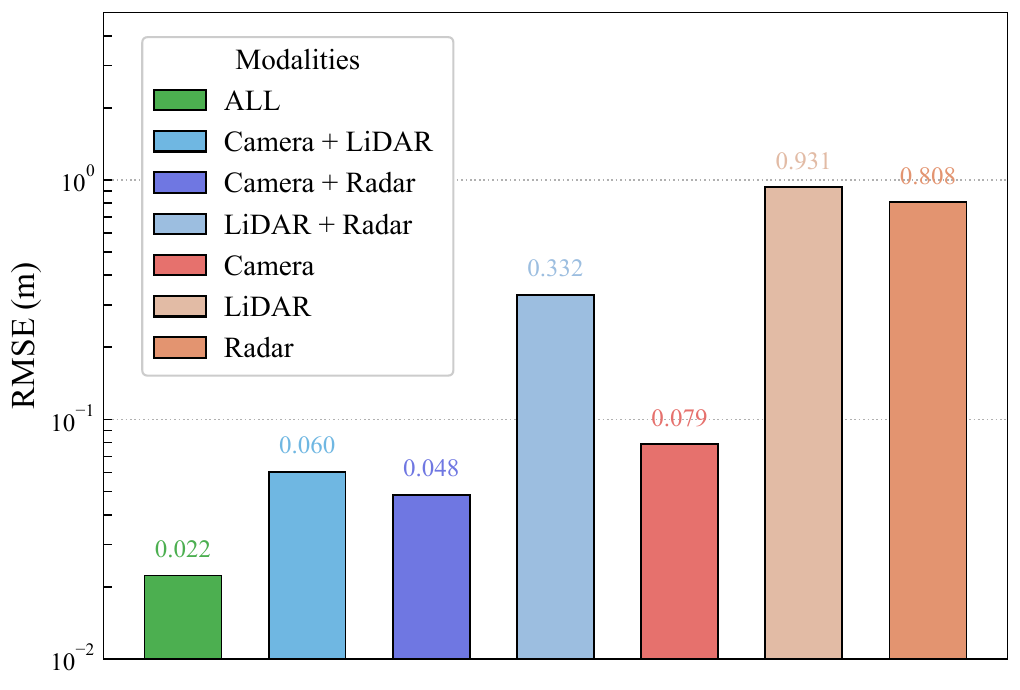}
    \captionsetup{skip=-1pt}
    \caption{Localization errors with pre-processing.}
    \label{all1}
  \end{subfigure}
  \medskip
  
  \begin{subfigure}{0.7\linewidth}
    \centering
    \includegraphics[width=\linewidth]{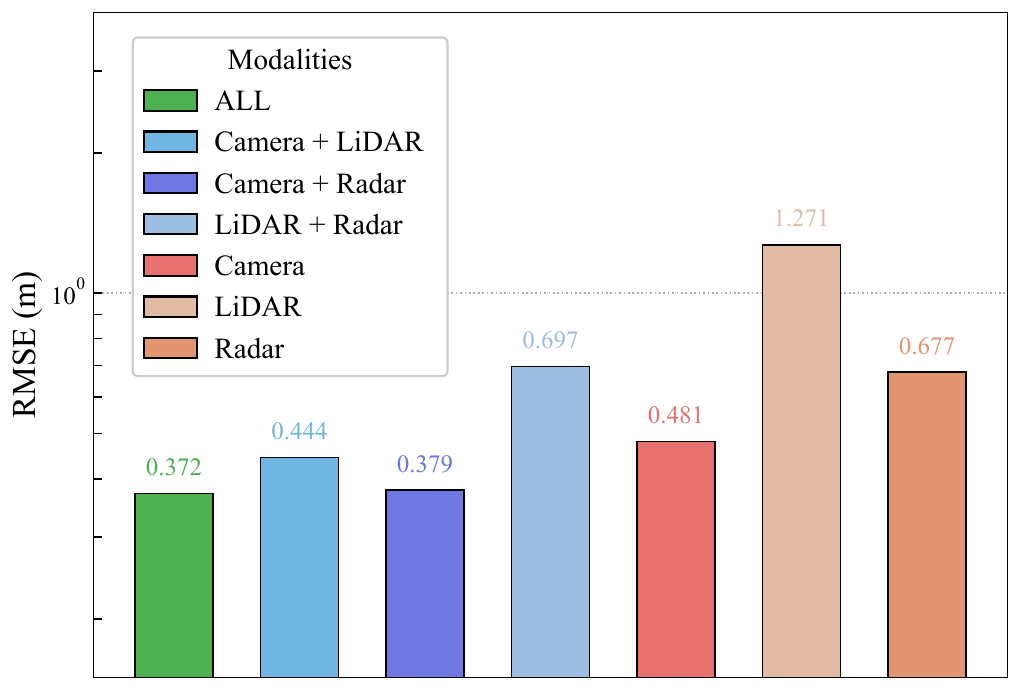}
    \captionsetup{skip=-1pt}
    \caption{Localization errors without pre-processing.}
    \label{all2}
  \end{subfigure}
  \caption{Localization errors under different combinations of sensing
    modalities.}
\end{figure}

\subsection{Performance of Complete Framework}

The performance of UAV localization of the proposed architecture in \fref{fig:mib} is evaluated and compared with the following baseline models: (i) a model directly processing the raw sensing data without any pre-processing, (ii) models employing Gated Recurrent Unit (GRU) \cite{dey2017gate}, Long Short-Term Memory (LSTM) \cite{hochreiter1997long}, and Temporal Convolutional Network (TCN) \cite{lea2017temporal} for temporal regression modeling assuming that the fusion module is unchanged, (iii) models employing different temporal sampling methods while keeping both the fusion module and the Mamba-based temporal regression network unchanged, and, (iv) models having the MADW and DACA fusion modules gradually added while keeping the pre-processing method, TE-Align, and Mamba network unchanged.

The localization errors are reported as RMSE values, and they are shown in \fref{all1} and \fref{all2}, respectively, assuming different modality combinations, and assuming that the data pre-processing is either included or excluded. In all cases, the multimodal fusion from \fref{fig:fusion} is used before the final temporal regression is performed by the Mamba network. From these results, it is clear that localization performance improves with the number of modalities included. This is true whether the pre-processing of sensing data has been used or not.

Furthermore, in order to evaluate the impact of camera pre-processing on the localization performance, we conducted additional experiments; these results are reported in \tref{pre_camera}. Thus, disabling camera pre-processing degrades the localization accuracy, whereas using only the pre-processing of camera images still improves the performance compared to the case without any pre-processing. These results further demonstrate the effectiveness of the proposed ROI-based camera pre-processing method.

\begin{table}[t]
  \centering
  \caption{The effect of the proposed camera pre-processing on the localization
    performance.}
  \label{pre_camera}
  \resizebox{\linewidth}{!}{
    \begin{tabular}{|c|c|c|c|c|}
      \hline Apply pre-processing & No modality & LiDAR and radar & Camera &
      Three modalities \\ \hline RMSE (m) & 0.372 & 0.095 & 0.035 & 0.022 \\
      \hline
    \end{tabular} }
\end{table}

\begin{figure}[t]
  \centering
  \begin{subfigure}{0.7\linewidth}
    \centering
    \includegraphics[width=\linewidth]{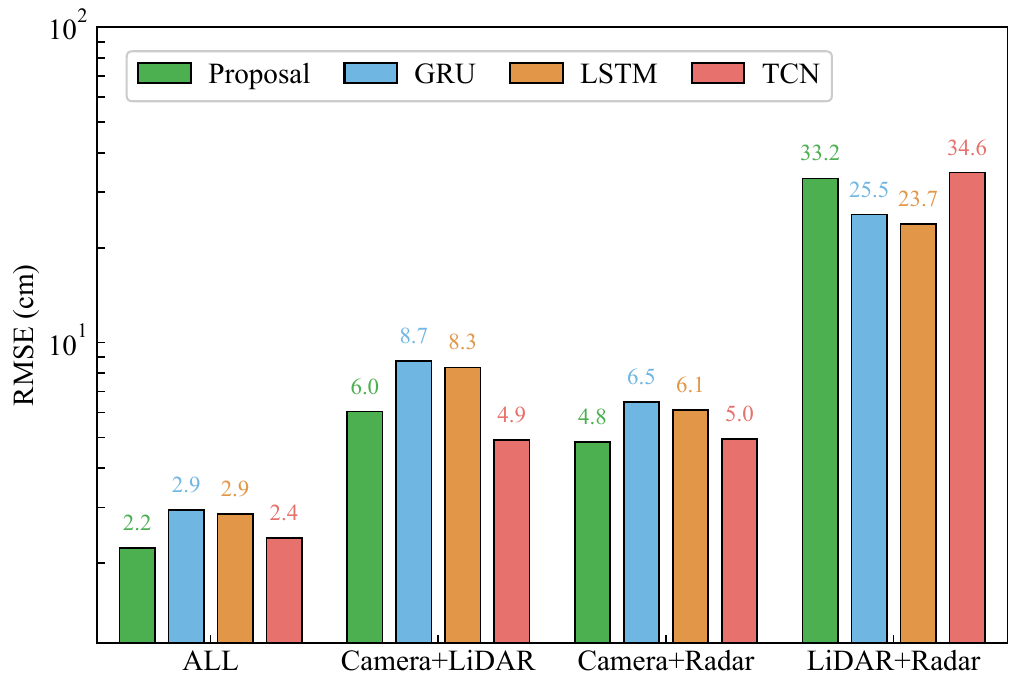}
    \captionsetup{skip=-1pt}
    \caption{Localization errors with pre-processing.}
    \label{all3_1}
  \end{subfigure}
  \medskip
  
  \begin{subfigure}{0.7\linewidth}
    \centering
    \includegraphics[width=\linewidth]{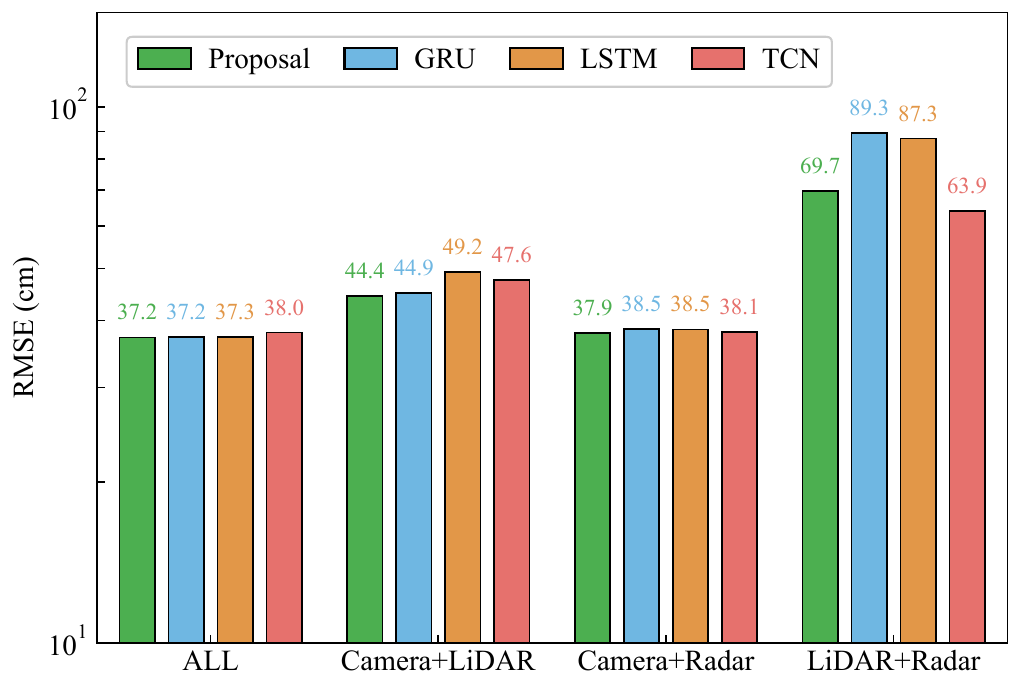}
    \captionsetup{skip=-1pt}
    \caption{Localization errors without pre-processing.}
    \label{all3_2}
  \end{subfigure}
  \caption{Localization errors under different temporal neural networks.}
  \label{all3}
\end{figure}

The localization performances of different temporal regression networks including Mamba, GRU, LSTM, and TCN are compared in \fref{all3_1} and \fref{all3_2}, assuming that the fusion module is kept unchanged. As shown in \fref{all3_1} with sensing data pre-processing, the proposed method maintains a competitive performance across different modal combinations, and it achieves the smallest RMSE, provided that all three sensing modalities are included. In contrast, as shown in \fref{all3_2}, directly feeding the raw sensing data into the localization network significantly increases localization errors for all temporal regression networks and also reduces their performance differences. These results demonstrate that the sensing data pre-processing plays a crucial role in improving the localization accuracy. In addition, Mamba is found to enhance the temporal regression performance using the same fusion module.

\begin{figure}[t]
  \centering
  \includegraphics[width=0.7\linewidth]{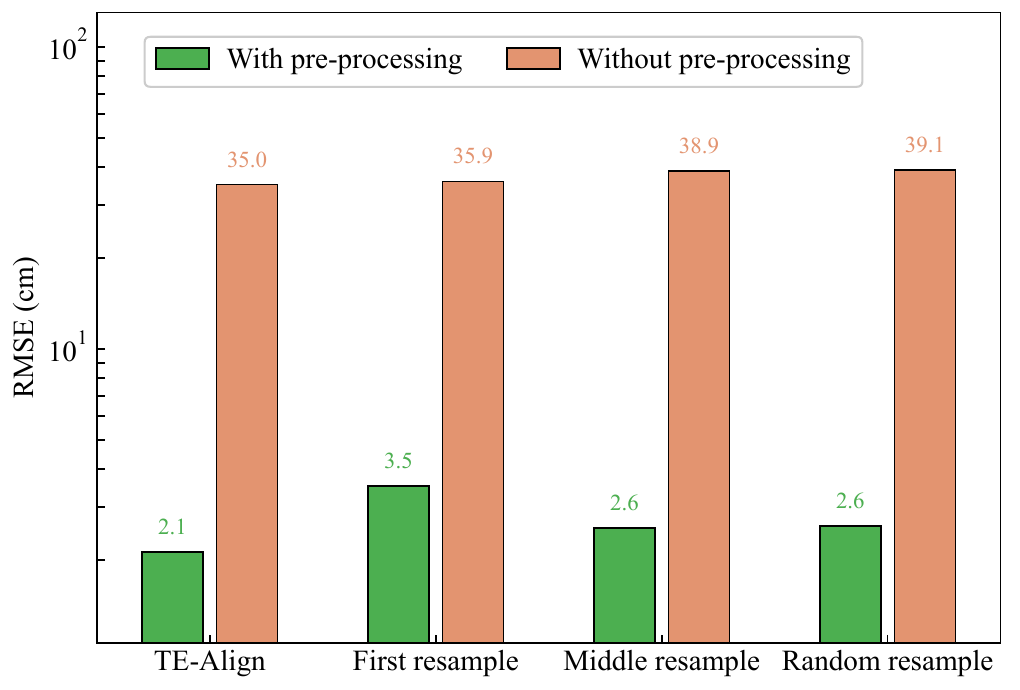} 
  \caption{Localization errors under different time alignment methods.}
  \label{TE}
\end{figure}

The localization performance under different temporal alignment methods is compared in \fref{TE}. The proposed TE-Align consistently outperforms the sampling-based methods, regardless of whether the data pre-processing is employed or not. The multimodal data are sampled within a $0.2$ \si{s} alignment window defined by the shared timestamps. The first-resample, middle-resample, and random-resample strategies, respectively, refer to selecting the first, the middle, and a randomly chosen sample from each modality within the alignment window. In order to suppress the influence of missing modalities, the random modality dropping is disabled during training. These results demonstrate that TE-Align can more effectively exploit the temporal correlations across modalities, and subsequently, improve the localization accuracy.

\begin{figure}[t]
  \centering
  \includegraphics[width=0.7\linewidth]{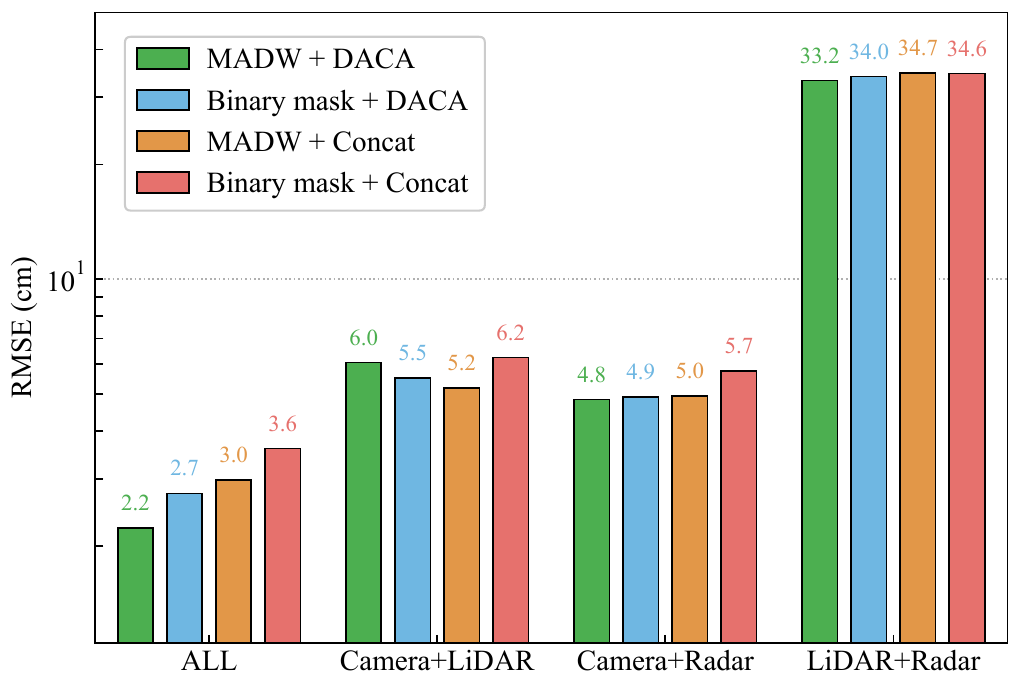}
  \caption{Localization errors for different fusion schemes.}
  \label{all4}
\end{figure}

The localization performance under different combinations of the fusion modules is compared in \fref{all4}. In most cases, the best performance is achieved when both MADW and DACA are enabled, which proves the effectiveness of combining the confidence-aware modality weighting with the cross-modal feature interactions. This improvement is particularly evident for modality combinations involving the radar data, possibly because the radar measurements are more susceptible to noise and occasionally missing frames. In such a case, MADW reduces the contributions of less informative modality features, while DACA extracts complementary information from available modalities. One exception to this rule can be observed for the camera-LiDAR combination, where enabling both modules leads to a slight performance degradation compared with using only MADW or DACA. The possible reason is that the pre-processed camera and LiDAR data already provide relatively stable and complementary information. In such a case, employing subsequent feature-weighting and interaction mechanisms is redundant, and it may even suppress certain modality-specific features. These results suggest that the effectiveness of the fusion modules depends on the quality and variations of the input sensing modalities.

In order to investigate the influence of temporal window size, $t_w$, the value of $t_w$ was varied from $0.1$ \si{s} to $0.5$ \si{s}, while keeping other settings unchanged. The corresponding localization errors are summarized in \tref{window}. It can be observed that, $t_w=0.2$ \si{s}, achieves the smallest localization error, and, $t_w=0.3$ \si{s}, has a comparable performance. When $t_w$ becomes larger, the increased temporal variations may introduce motion-induced ambiguity. At the same time, the excessively small window contains fewer observations due to heterogeneous sampling rates, which leads to less stable feature aggregation. Consequently, the value, $t_w=0.2$ \si{s}, appears to provide a good trade-off between the temporal coverage and the motion-induced drift.

\begin{table}[t]
  \centering
  \caption{Analysis of the influence of temporal window length.}
  \label{window}
  \resizebox{0.8\linewidth}{!}{
    \begin{tabular}{|c|c|c|c|c|c|}
      \hline
      Window $t_w$ (s) & 0.1 & 0.2 & 0.3 & 0.4 & 0.5\\
      \hline
      RMSE (cm) &2.997& 2.223& 2.254& 2.467& 2.496\\
      \hline
    \end{tabular} }
\end{table}

\begin{figure}[t]
  \centering
  \begin{subfigure}{0.7\linewidth}
    \centering
    \includegraphics[width=\linewidth]{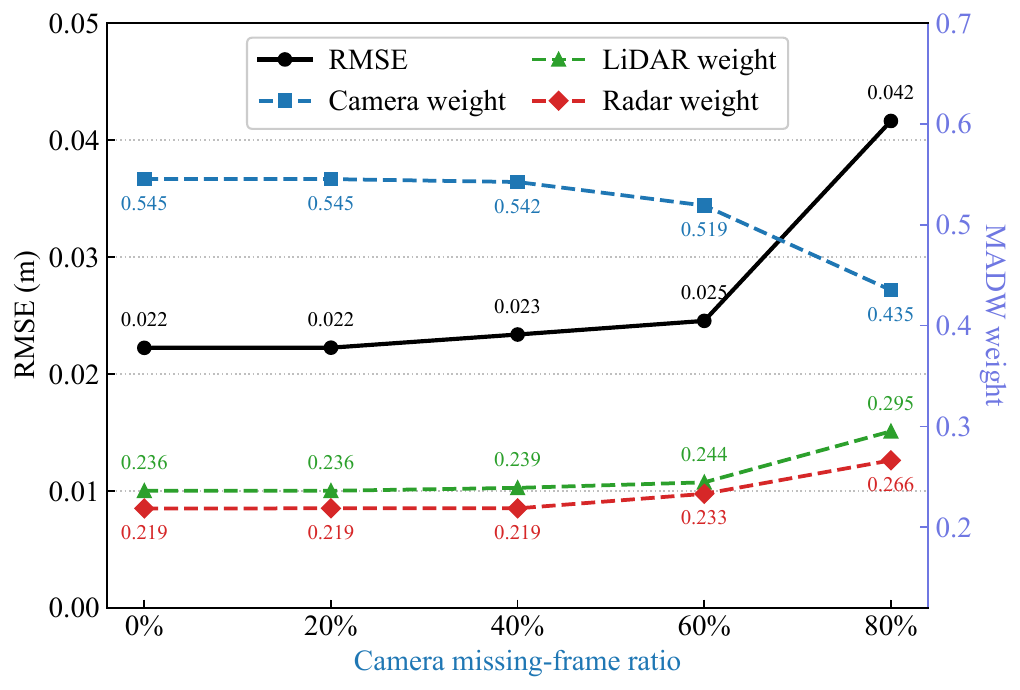}
    \captionsetup{skip=-1pt}
    \caption{The effect of camera data degradation.}
    \label{R2C}
  \end{subfigure}
  \medskip
  
  \begin{subfigure}{0.7\linewidth}
    \centering
    \includegraphics[width=\linewidth]{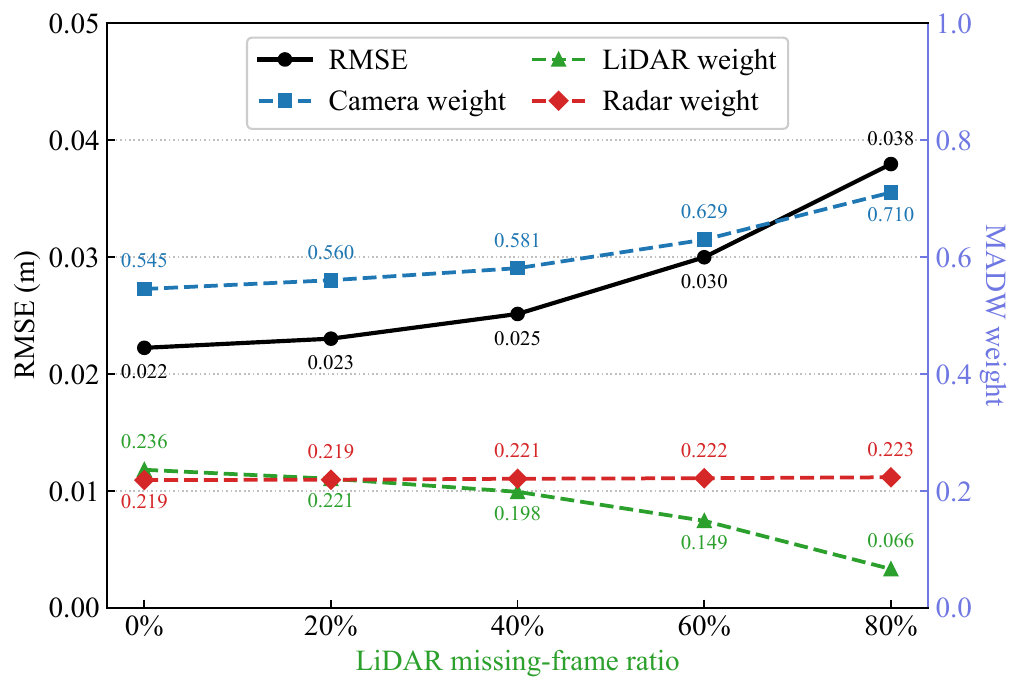}
    \captionsetup{skip=-1pt}
    \caption{The effect of LiDAR data degradation.}
    \label{R2L}
  \end{subfigure}
  \medskip
  
  \begin{subfigure}{0.7\linewidth}
    \centering
    \includegraphics[width=\linewidth]{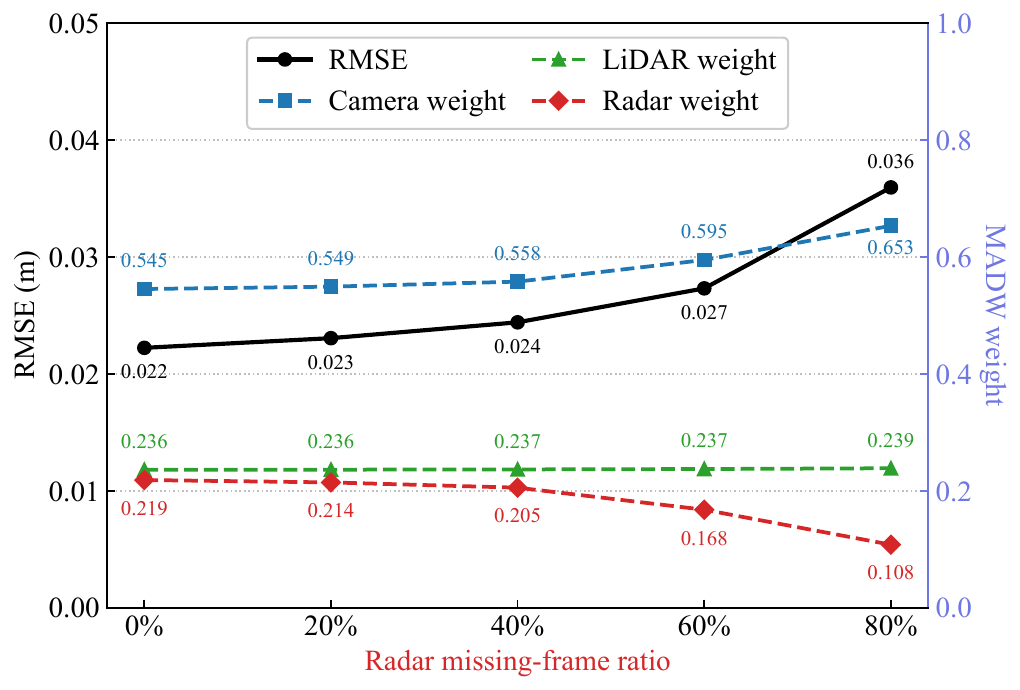}
    \captionsetup{skip=-1pt}
    \caption{The effect of radar data degradation.}
    \label{R2R}
  \end{subfigure}
  \caption{The effect of modality degradation on the localization error and the
    MADW weights.}
  \label{MADW}
\end{figure}

The robustness of MADW was further evaluated under gradually degraded sensing conditions. In particular, the missing frames were progressively introduced into incoming data streams. The missing-frame ratio was varied from 0\% to 80\% separately for each modality, and the localization error and the learned modality weights were analyzed. The missing frames were introduced only during the model evaluation, but not during the model training. The results are shown in \fref{MADW}. It can be observed that when the missing-frame ratio of a given modality increases, the corresponding MADW weight gradually decreases, and the weights of other modalities increase in order to compensate for degraded information. The adaptive re-weighting demonstrates that MADW can follow varying informativeness of the latent modality representations under dynamically changing sensing conditions.

\begin{figure}[t]
  \centering
  \includegraphics[width=0.7\linewidth]{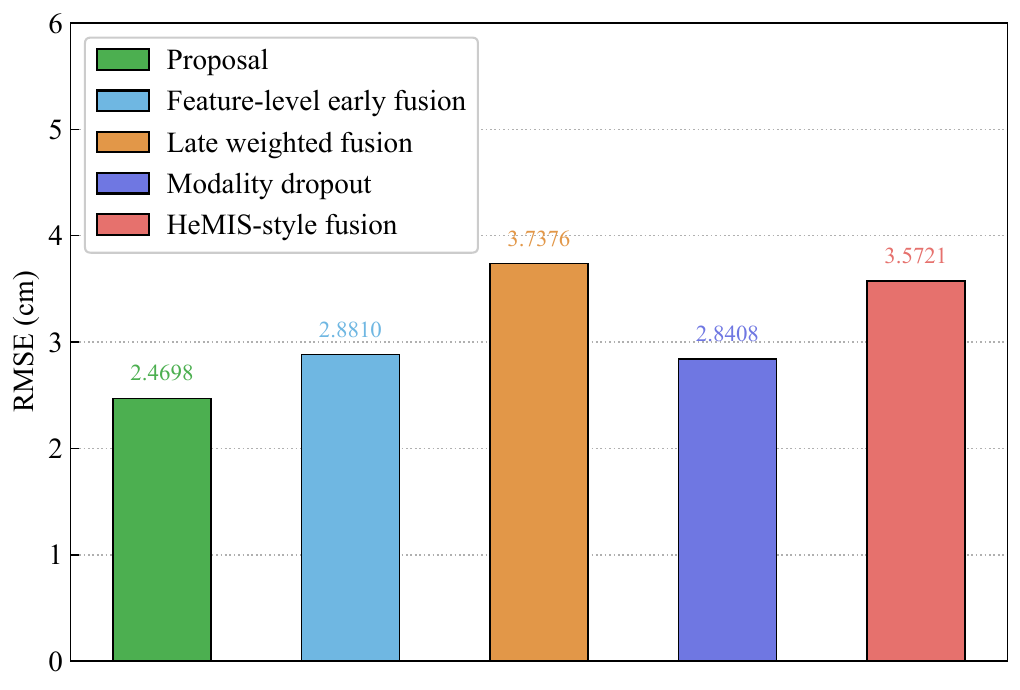} 
  \caption{Localization errors for different multimodal fusion methods.}
  \label{sys}
\end{figure}

Finally, the performance of the proposed framework was compared with several other representative multimodal fusion baselines. Since the existing multimodal fusion localization methods are not directly applicable to the scenario and the dataset considered in this paper, the representative baselines had to be adapted assuming their core fusion principle. The following fusion methods were considered: the feature-level early fusion \cite{liu2023bevfusion}, the late weighted fusion \cite{ali2026cross}, the modality dropout \cite{neverova2015moddrop, park2025resilient}, and the HeMIS-style fusion \cite{havaei2016hemis, wang2023multi}. In particular, the feature-level early fusion projects the features of three modalities into a shared feature space, concatenates them, and then performs the temporal modeling. The late weighted fusion first predicts the target position using separate unimodal branches, and then performs a decision-level fusion with learnable weights. The modality dropout is built upon an early fusion. It randomly drops some modalities during training to improve the robustness against the missing and degraded modalities. The HeMIS-style fusion maps each modality into a shared feature space, and then aggregates the available features using their mean and variance, which provides a mechanism for adapting to missing and degraded modalities. The localization errors of these methods are compared in \fref{sys}. It can be observed that the proposed method outperforms all other baseline methods considered.

\subsection{Communication Application}

\begin{figure}[t]
  \centering
  \includegraphics[width=0.7\linewidth]{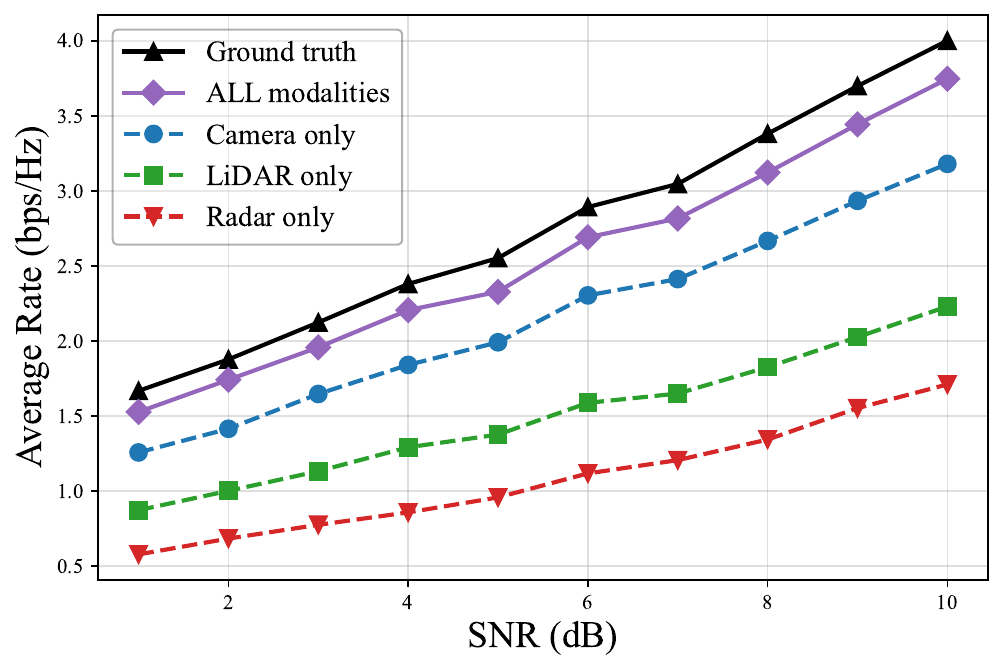}
  \caption{Achievable data rates as a function of SNR.}
  \label{comm}
\end{figure}

The UAV can serve as a communication relay to extend the BS coverage and the achievable data rates. Thus, assume that the BS is equipped with an $M^x \times M^y$ two-dimensional UPA, and one antenna is mounted on the UAV. The mmWave channel with $P$ paths in the downlink can be modeled as,
\begin{equation}
  \bm{h} = \sum_{p=1}^P \alpha_p \bm{a}(\theta_p,\phi_p) \in
  \mathbb{C}^{(M^x M^y \times 1 )},
\end{equation}
where $\alpha_p$ is the complex-valued gain of the $p$-th path, $\bm{a}(\theta_p,\phi_p)$ is the array steering vector, $ \theta$ denotes the azimuth, and $\phi$ denotes the elevation \cite{he2023integrated}. Given the present UAV position, $(x_{\text{est}}, y_{\text{est}}, z_{\text{est}})$, which can be inferred by one of the sensing schemes discussed above, the BS steers the unit-norm beamforming vector, $\bm{f}_{\text{est}}$, in the direction $ (\theta_{\text{est}}, \phi_{\text{est}} )$, which yields the effective channel gain, $\gamma = | \bm{h}^H \bm{f}_{\text{est}}|^2$. The corresponding data rates as a function of the signal-to-noise ratio (SNR) are reported in \fref{comm}. It can be observed that the multimodal fusion is useful not only for improving the UAV target localization, but it can also greatly assist in improving the achievable data rates in communication networks. \fref{comm} also shows that the achievable data rates with inferred UAV location using all three modalities are approaching the data rates when the UAV location is known exactly.

\section{Conclusion}  \label{sc:7}

A multimodal UAV localization scheme referred to as \name has been designed and analyzed. This scheme is intended for a collaborative deployment on the resource-constrained sensing nodes, and the edge server attached to a BS. The sensing nodes perform modality-specific feature extraction and the IB-based compression in order to map the features to a shared latent space. This alleviates the transmission requirements when offloading the data to the edge server. The edge server performs temporal alignment and multimodal fusion using the designed TE-Align, MADW, and DACA modules. The subsequent inference of the UAV location is done using a Mamba-based model, which enables accurate long-term predictions of the target locations. Extensive numerical experiments were carried out using the real-world dataset. The UAV localization performance was evaluated assuming different system configurations. These results showed that the devised data pre-processing has a strong positive influence on localization accuracy. The best performance was always achieved when all three modalities are considered. In good visibility conditions, camera sensing achieves the best performance, and LiDAR sensing outperforms radar sensing. However, the performance of different sensing modalities is strongly dependent on the environmental conditions, and the availability of good-quality sensing data.

Future work can extend the localization framework to multiple concurrent targets, possibly considering both aerial and ground targets. The localization framework could be expanded to enable continuous collaborative monitoring of much larger geographical areas when the sensing data are shared across a communication network among many BSs.

\bibliographystyle{IEEEtran}
\bibliography{reference.bib}

\begin{IEEEbiography}
  [{\includegraphics[width=1in,clip,keepaspectratio]{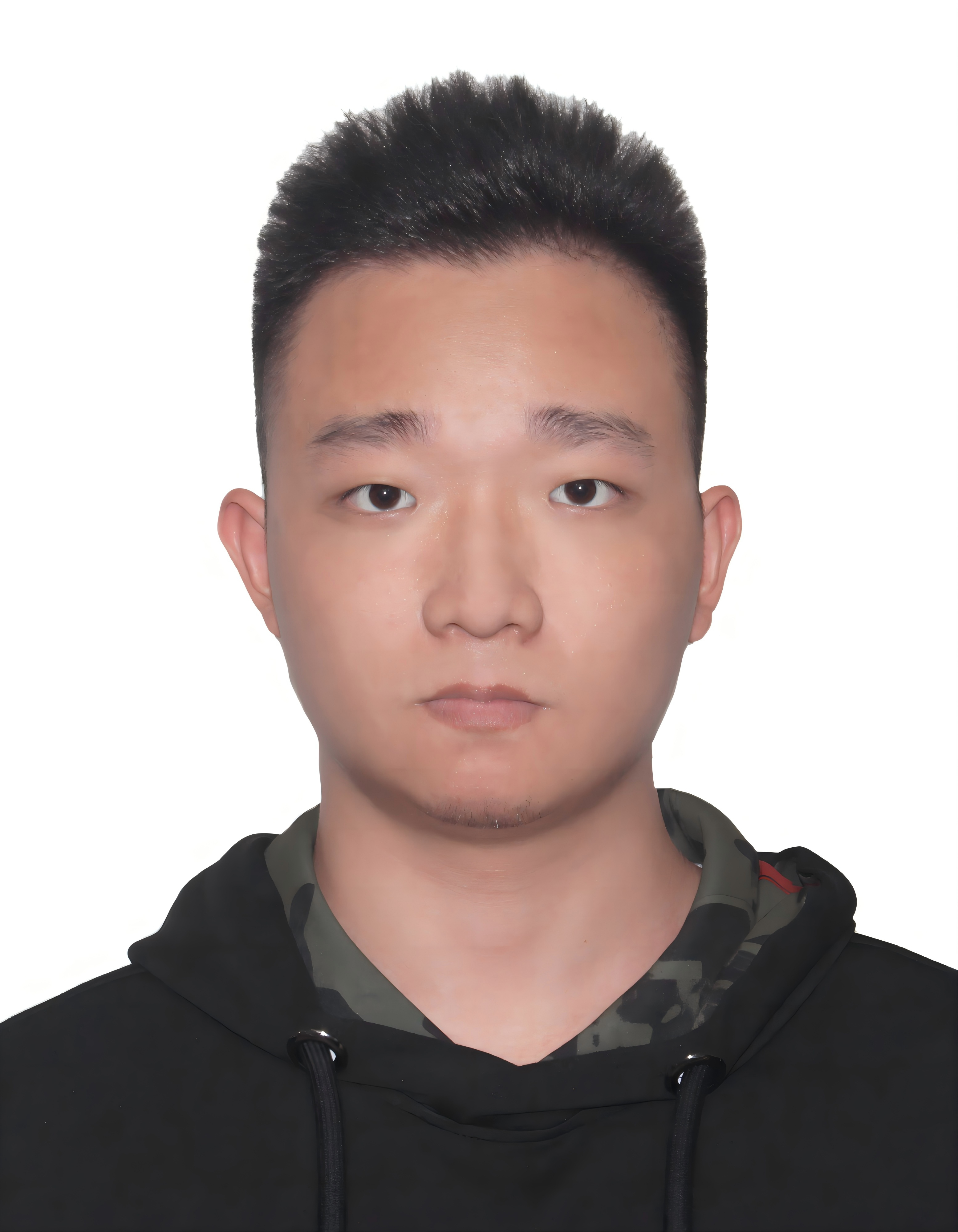}}] {Zhong Ye} received the B.E. degree in communication engineering from Zhejiang University City College, Hangzhou, China, in 2019, and the M.Sc. degree in information and communication engineering from Zhejiang Gongshang University, Hangzhou, China. He is currently pursuing the Ph.D. degree at Zhejiang University, and his research interests mainly include vehicular networking systems, multimodal sensing fusion, and integrated sensing and communications (ISAC).
\end{IEEEbiography}

\begin{IEEEbiography}
  [{\includegraphics[width=1in,clip,keepaspectratio]{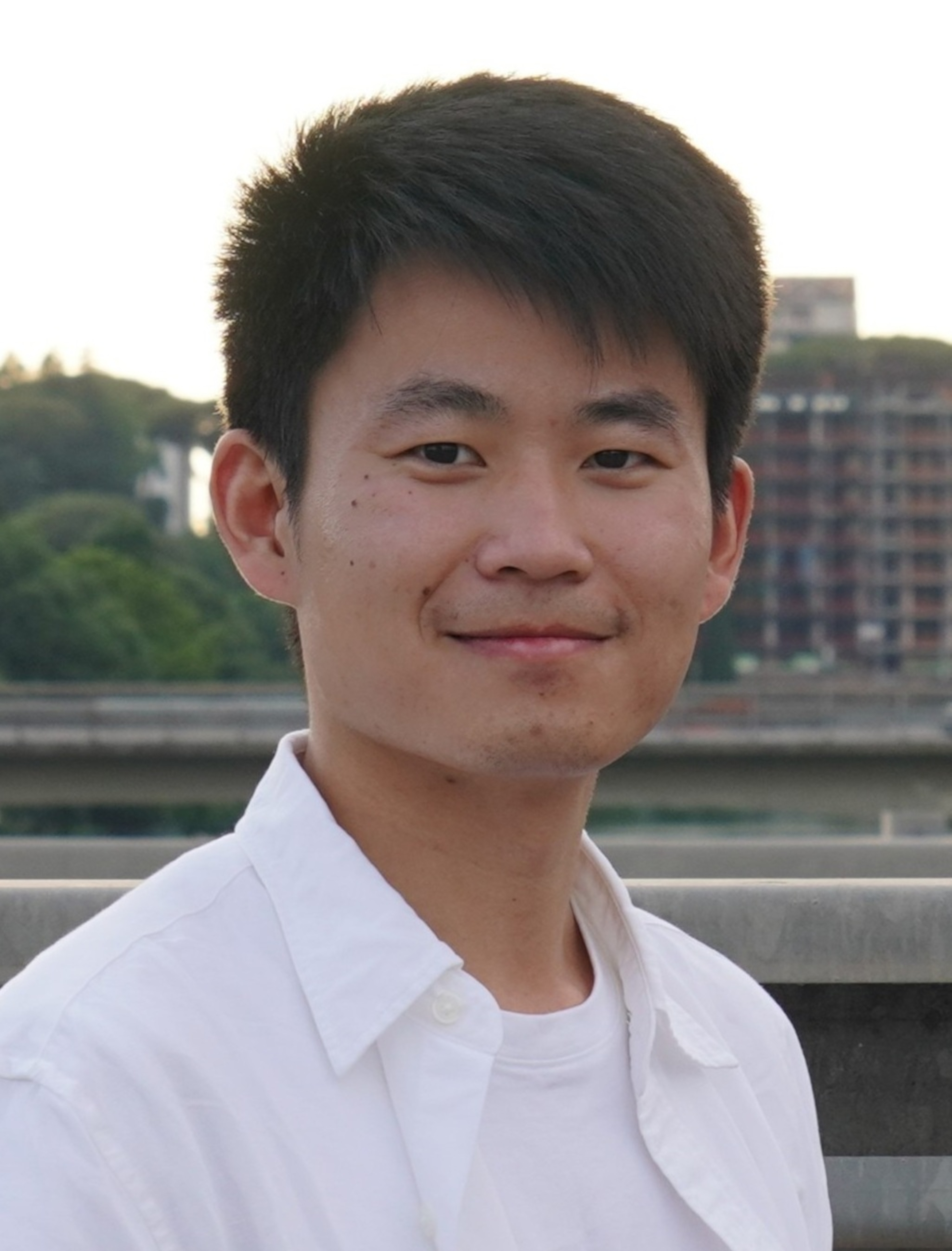}}] {Yinghui He} (Member, IEEE) received the B.E. degree in information engineering and Ph.D. degree in information and communication engineering from Zhejiang University, Hangzhou, China, in 2018 and 2023, respectively. 
  His research interests mainly include integrated sensing and communications (ISAC), mobile computing, and device-to-device communications.
\end{IEEEbiography}

\begin{IEEEbiography}
  [{\includegraphics[width=1in,clip,keepaspectratio]{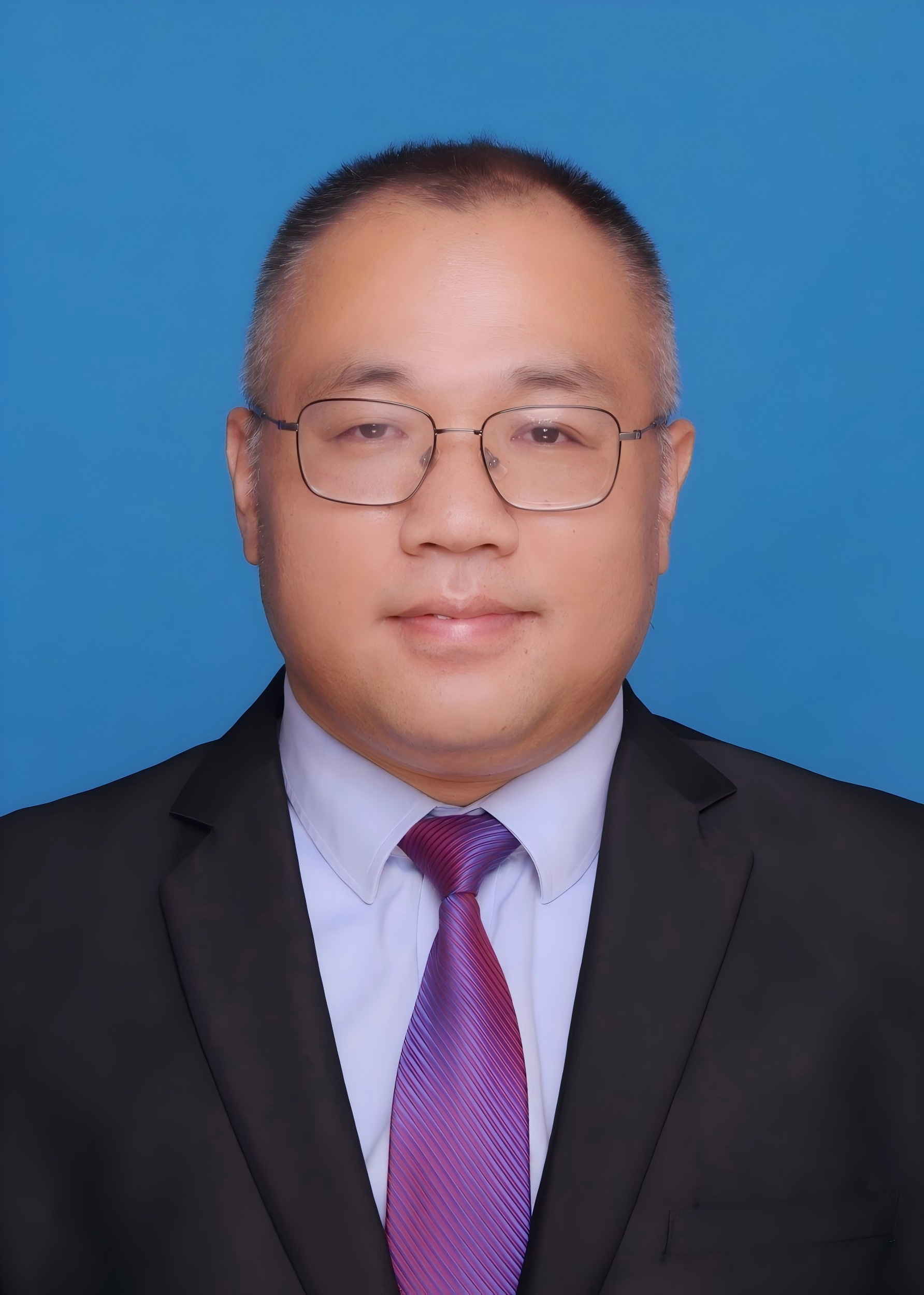}}]
  {Guanding Yu} (Senior Member, IEEE) received the B.E. and Ph.D. degrees in communication engineering from Zhejiang University, Hangzhou, China, in 2001 and 2006, respectively. He joined Zhejiang University in 2006, and is now a Professor with the College of Information and Electronic Engineering. From 2013 to 2015, he was also a Visiting Professor at the School of Electrical and Computer Engineering, Georgia Institute of Technology, Atlanta, GA, USA. His research interests include integrated sensing and communications (ISAC), mobile edge computing/learning, and machine learning for wireless networks.

  Dr. Yu has served as a guest editor of IEEE Communications Magazine special issue on Full-Duplex Communications, an Editor of IEEE Journal on Selected Areas in Communications Series on Green Communications and Networking, and Series on Machine Learning in Communications and Networks, an Editor of IEEE Wireless Communications Letters, a lead Guest Editor of IEEE Wireless Communications Magazine special issue on LTE in Unlicensed Spectrum, an Editor of IEEE Transactions on Green Communications and Networking, and an Editor of IEEE Access. He is now serving as an editor of \emph{IEEE Transactions on Machine Learning in Communications and Networking}. He received the IEEE ComSoc Asia-Pacific Outstanding Young Researcher Award in 2016 and the IEEE ComSoc Stephen O. Rice Prize in 2026. He regularly sits on the technical program committee (TPC) boards of prominent IEEE conferences such as ICC, GLOBECOM, and VTC. He also serves as a Symposium Co-Chair for IEEE GLOBECOM 2019 and a Track Chair for IEEE VTC 2019'Fall.
\end{IEEEbiography}

\begin{IEEEbiography}
  [{\includegraphics[width=1in,clip,keepaspectratio]{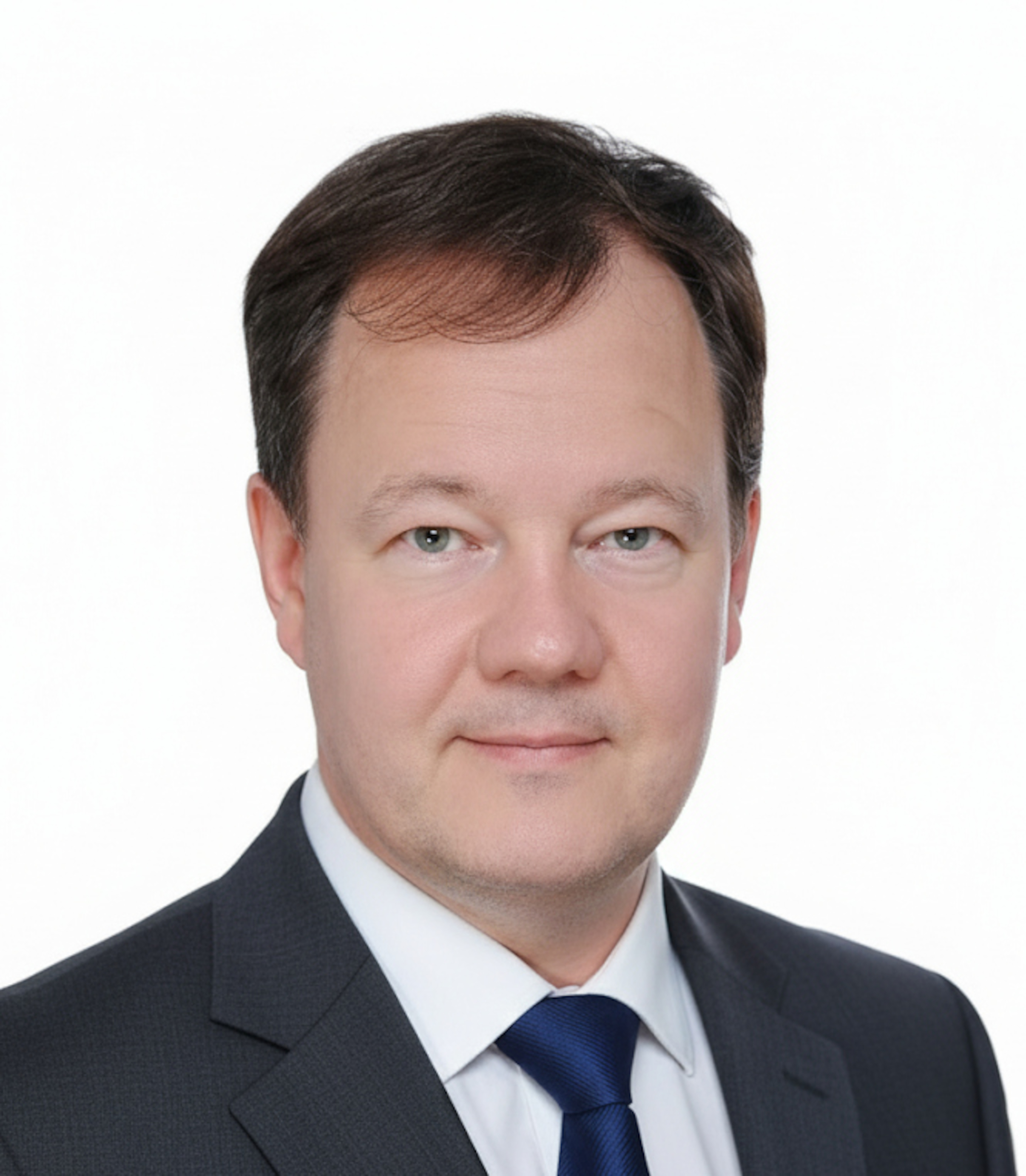}}] {Pavel Loskot} (Senior Member, IEEE) received the B.Sc. and the M.Sc. degrees in biomedical electronics and radio electronics, respectively, from the Czech Technical University of Prague, Czech Republic, and the Ph.D. degree in wireless communications from the University of Alberta, Canada. Before joining ZJU-UIUC Institute as Associate Professor in 2021, he was a Senior Lecturer at Swansea University, UK. In academic year 2014/2015, he was a visiting researcher at Computational Science Research Center, Beijing, China. From 1999 to 2001, he was a Research Scientist at the Centre for Wireless Communications, Oulu, Finland. His current research interests focus on mathematical modeling, statistical signal processing and machine learning for multi-sensor and time-series data.
	
  Prof. Loskot is the Elected IARIA Fellow 2025, Fellow of the Higher Education Academy, UK, and holds the Recognized Research Supervisor distinction by the UK Council for Graduate Education. He is a technical committee member of many IEEE conferences annually. From 2014 till 2020, he served on the IEEE Membership Development Team and Selection Committee of the UK and Ireland Section.
\end{IEEEbiography}
\vfill

\end{document}